\documentclass[aps,prb,preprint,showpacs,superscriptaddress,floatfix]{revtex4}
\usepackage{graphics}
\usepackage{epsfig}
\usepackage{amsmath}
\usepackage{amssymb}
\usepackage{calc}
\newcommand{\beq}{\begin{equation}}
\newcommand{\eeq}{\end{equation}}
\newcommand{\bea}{\begin{eqnarray}}
\newcommand{\eea}{\end{eqnarray}}

\begin{document}
\title{Two-Center Black Holes, Qubits and Elliptic Curves}
\author{P\'eter L\'evay}
\affiliation{Department of Theoretical Physics, Institute of
Physics, Budapest University of Technology, H-1521 Budapest,
Hungary}
\date{\today}
\begin{abstract}

We relate the U-duality invariants characterizing
two-center extremal black hole solutions in the $stu$, $st^2$ and
$t^3$ models of $N=2$, $d=4$ supergravity  to the basic invariants
used to characterize entanglement classes of four-qubit systems.
For the elementary example of a D0D4-D2D6 composite in the $t^3$
model we illustrate how these entanglement invariants are related
to some of the physical properties of the two-center solution.
Next we show that it is possible to associate elliptic curves to
charge configurations of two-center composites. The
hyperdeterminant of the hypercube, a four-qubit polynomial
invariant of order $24$ with $2894276$ terms, is featuring the
$j$ invariant of the elliptic curve. We present some evidence that
this quantity and its straightforward generalization should play
an important role in the physics of two-center solutions.
\end{abstract}
\maketitle{}

\section{Introduction}

The aim of the present paper is to show that it is possible to
relate the entanglement measures usually used in studies
concerning four-qubit systems to the U-duality invariants found
recently by Ferrara et. al.\cite{Ferretal} characterizing extremal
two-center black hole solutions in the $stu$, $st^2$ and $t^3$
models. Interestingly as a byproduct of establishing this
correspondence one can also come across an interesting connection
between charge configurations of such black holes and a special
class of elliptic curves.

Multi center black hole solutions provide an interesting research
direction within the rapidly evolving field of black hole
solutions in supergravity, string and M-theory. For such solutions
the attractor mechanism\cite{attractor} has been generalized
giving rise to split attractor flows\cite{Denef2,Bates}. For
two-center solutions the latter term refers to the situation, when
in moduli space after crossing walls of marginal stability the
attractor flows are separately evolving to the attractor points of
the constituent single center solutions. Recently these
developments have triggered activity in a variety of new research
fields such as attractor flow trees, entropy enigmas, microstate
counting, and bound state
recombination\cite{Mooredenef,Castro,Sen,Sen2,David,Verlinde}.

For single center solutions it has become obvious that the notion
of duality charge orbits and their invariants\cite{orbits} are
useful concepts for classifying black hole solutions together with
their supersymmetry properties. Except for a special
case\cite{exception} for multicenter solutions the corresponding
structure of orbits and invariants is still unknown. In order the
remedy this situation in a recent paper Ferrara et.al.
conducted\cite{Ferretal,Ferretal2} a systematic study on the
structure of invariants characterizing the charge configurations
and invariants of two center solutions. In the case of the $stu$,
$st^2$ and $t^3$ models looking at the structure of such
invariants one immediately notices structural similarities to the
well known sets of four-qubit invariants\cite{Luque,Levay4}
discussed in the seemingly unrelated field of quantum information.

 Since the advent of the black
hole qubit correspondence\cite{BHA1} such coincidences should not
come as a surprise. It is well-known by now that few qubit
entangled systems are capable of providing interesting new insight
into the structure of black hole solutions and their attractor
flows\cite{Levay3}. The occurrence of these qubits is related to
the presence of tensor
 products of the spin $\frac{1}{2}$  irreducible representations
of $SL(2)$ groups. Such products of $SL(2)$s show up as subgroups
of $U$-duality groups governing the entangled web of dualities of
supergravity models giving rise to black hole solutions. Initially
mathematical coincidences were recorded merely for two and three
qubit systems and the corresponding axion-dilaton and $stu$ black
holes\cite{BHA1}, however evidence for $n$-qubit systems with $n>3$ to
make their presence in this context started to accumulate.

In this line of development the relevance of four qubit systems to
black hole solutions in supergravity was first pointed out in our paper
\cite{fanolevay} where the isomorphism
\beq so(4,4)\simeq
sl(2)^4\oplus(2,2,2,2)
\label{isom}\eeq\noindent has been used to
describe $7\times 16$  of the $133$ $E_{7(7)}$ generators of
$N=8$, $d=4$ supergravity. These generators describing seven copies of four-qubit states are not belonging to
the $sl(2)^7$ subalgebra of $E_{7(7)}$. A suitable incidence
geometry accounting for the relationship between the $7$ groups of $16$
generators is that of the dual Fano plane giving rise to a
geometry, dual to the one describing the "tripartite entanglement
of seven qubits" interpretation\cite{ferfano,fanolevay} of the
quartic $E_{7(7)}$ black hole entropy formula. The isomorphism of
Eq.(\ref{isom}) was also discussed in the review paper of Borsten
et. al.\cite{Borsten3} providing further interesting examples of
simple qubit systems.

In our next paper\cite{Levayfour} in a four-qubit entanglement based
formalism the structure of extremal stationary spherically
symmetric black-hole solutions in the STU model of $N=2$, $d=4$
supergravity was described. The basic idea facilitating this
interpretation was the fact that stationary solutions in $d=4$
supergravity can be described by dimensional reduction along the
time direction\cite{Breitenlohner}. In this $d=3$ picture the global
symmetry group $SL(2,{\mathbb R})^{×3}$ of the STU model is
extended by the Ehlers $SL(2,{\mathbb R})$ accounting for the
fourth qubit. One can then introduce a four-qubit state depending
on the charges, the moduli, and the warp factor. Here it was also
noticed that in the terminology of four-qubit entanglement
extremal black-hole solutions should correspond to nilpotent, and
nonextremal ones to semisimple states. The upshot of these
considerations was the emerging possibility of relating the
entanglement properties of such and similar states to different
classes of black-hole solutions in the STU model. The challenge of
elaborating on this idea was recently taken up in the papers of
Borsten et.al.\cite{Borstenfour} In these papers the authors
applied the black-hole qubit correspondence to the problem of
classifying four-qubit entanglement. The key technical ingredient
was the Kostant-Sekiguchi theorem which establishes the link
between nilpotent orbits of extremal black holes and four qubit
entanglement types. The emerging picture is: we have $31$
entanglement families which reduce to nine up to permutations of
the qubits. These nice papers confirmed once again that the input
coming from string theory can be useful in establishing results in
a different field, since the literature until now on four qubit
entanglement classification was confusing and seemingly
contradictory.

In this paper we would like to show that the charge orbit
classification of two-center black hole solutions in the $stu$
model is another arena where four-qubit systems naturally
make their appearance. As a first possible step in this direction
here we establish a correspondence between the U-duality
invariants of Ferrara et.al.\cite{Ferretal} and the four-qubit
invariants showing up in classification schemes of entanglement
types in quantum information. Establishing this correspondence
simplifies some of the invariants proposed so far, clarifies their
geometric and algebraic roles, and provides hints for further
generalizations outside the framework of $N=2$, $d=4$
supergravity. As an extra bonus the four-qubit picture also hints
at a basic physical role these invariants are playing in the
theory of two-center solutions. For one of the invariants not
fully appreciated yet, our considerations establish a special role.
It is the $SL(2)^{\times 4}$ and permutation invariant
hyperdeterminant of type $2\times 2\times 2\times 2$. This is a
polynomial of order $24$ in the $16$ amplitudes of the  four-qubit
state. Mapping the $16$ amplitudes to the $16$ charges
characterizing two-center solutions in the $stu$ model, for a
special case we show that the structure of this hyperdeterminant
seem to govern issues of consistency in the realm of two-center
solutions. These ideas also suggest a natural way for associating
an elliptic curve of a special kind to a particular charge
configuration. The coefficients of our elliptic curve are the
algebraically independent four-qubit invariants, and its
discriminant is just the hyperdeterminant. We also present some
evidence for the conjecture that the structure of the $j$
invariant of the elliptic curve should play an important role in
the physical properties of the two-center solution.
The idea that elliptic functions and the $j$ invariant might play some role
in four-qubit systems and the black hole qubit correspondence
was first suggested by P. Gibbs\cite{Gibbs} some related discussion appeared in
the paper of Bellucci et.al.\cite{Bellucci}.

The organization of this paper is as follows. In Section II. we
summarize the background material on four-qubit invariants,
reduced density matrices, and the structure of the
hyperdeterminant of the hypercube. We introduce a quartic
polynomial featuring the algebraically independent four-qubit
$SL(2,{\mathbb C})$ invariants. In Section III. we are discussing
extremal two center black hole charge states in a four-qubit based
picture. Here we work out a dictionary between the invariants
found by Ferrara et. al.\cite{Ferretal} in the so called
Calabi-Visentini basis and the algebraically independent
four-qubit ones in the "special coordinates" basis. We give some
of the invariants used in Ref.\cite{Ferretal} a simpler
appearance, and connect other invariants of physical meaning to
properties of four-qubit reduced density matrices. Here we also
show that the set of algebraically independent polynomial
invariants in both the four-qubit and the Ferrara et.al.
description are based on two seemingly different quartic
polynomials however, with the {\it same} resolvent cubic.

Section IV. is devoted to a case study featuring BPS $D0D4-D2D6$
composites in the $t^3$ model. In the paper of Bates and
Denef\cite{Bates} this elementary example has already turned out
to be a good playing ground for investigating the basic properties
of two center solutions, hence we opted for illustrating the
physical role of our four qubit invariants in the very same
setting. These considerations relate the (necessary) consistency
condition, guaranteeing the BPS composite to exist, to the
positivity of the hyperdeterminant and to the extra constraint
that the two nonzero invariants of the $t^3$ model are having the
{\it same sign}. It turns out that precisely these conditions are
the ones guaranteeing the fundamental quartic polynomial to have
real roots. We then associate an elliptic curve of Weierstrass
canonical form to the resolvent cubic of this quartic and show how
physical properties are nicely encapsulated in the structure of
its $j$ invariant.

In Section V. we examine the status of our rather ad hoc
assignment: two-center charge configuration-elliptic curve, more
thoroughly. By switching to the most general Tate form of an
elliptic curve we show that our association of elliptic curves to
two-center black hole charge configurations in the $stu$ model is
a natural one. This means that the nonzero coefficients $a_j$ with
$j=1,2,3,4$ appearing in the Tate form are algebraically
independent four-qubit invariant homogeneous polynomials of order
$2j$. These coefficients have important physical meaning: $a_1$ is
just the canonical symplectic pairing between the charge vectors
of the two centers. Vanishing of $a_4$ gives rise to the $st^2$,
and a further vanishing of $a_2$ results in the $t^3$ truncation.
One can also see that in the $st^2$ and $t^3$ models sending $a_3$
to zero corresponds to the limit when our elliptic curve
degenerates. One of the nontrivial coefficients $a_6$ in the Tate
form is always zero for the $stu$ model. This is related to the
vanishing of a nontrivial polynomial constraint of homogeneous
degree $12$ valid in the $stu$ model, already observed by Ferrara
et.al.\cite{Ferretal} Based on the latest results of
Andrianopoli et. al.\cite{Ferretal2} we conjecture that we should be able to
generalize our correspondence between charge orbits and elliptic
curves also for the case of maximal $N=8$, $d=4$ supergravity. In
this case the $stu$ model should arise as an $a_6=0$ truncation
implemented by the vanishing of a polynomial of order $12$.

The aim of our last speculative Section VI. is to draw the readers
attention to some interesting structural similarities showing up
in a variety of physical contexts where our four-qubit invariants
parametrizing elliptic curves might play a crucial role. Here we
give a new look and interpretation to a triality invariant curve
originally introduced by Seiberg and Witten\cite{SW}. Now this
curve is paramaterized by four-qubit invariants also displaying
permutation invariance. In this new setting we also invoke the
F-theory interpretation of this curve as was given by
Sen\cite{senorient}. Finally our conclusions  and some comments
are left for Section VI.

\section{Four qubit systems}

In order to facilitate a four-qubit description of the two center
charge configurations in the $stu$ model our aim in this
subsection is to review the background material on four-qubit
states and their entanglement measures. A four qubit state can be
written in the form

\beq \vert {\Lambda}\rangle
=\sum_{i_0i_1i_2i_3=0,1}{\Lambda}_{i_0i_1i_2i_3}\vert
i_0i_1i_2i_3\rangle,\quad \vert i_0i_1i_2i_3\rangle\equiv \vert
i_0\rangle\otimes\vert i_1\rangle\otimes\vert i_2\rangle\otimes
\vert i_3\rangle\in V_0\otimes V_1\otimes V_2 \otimes
V_3
\label{Lambda} \eeq \noindent where $V_{0,1,2,3}\equiv {\mathbb
C}^2$. Let the subgroup of stochastic local operations and
classical communication\cite{Dur} representing admissible
fourpartite manipulations on the qubits be just $SL(2, {\mathbb C})^{\otimes 4}$ acting
on $\vert\Lambda\rangle$ as

\beq \vert\Lambda\rangle\mapsto (S_0\otimes S_1\otimes S_2\otimes
S_3)                  \vert\Lambda\rangle,   \quad S_{\alpha}\in
SL(2, {\mathbb C} ) ,\quad\alpha=0,1,2,3. \eeq \noindent Our aim
is to give a unified description of four-qubit
states taken together with their SLOCC transformations and their
associated invariants. As we will see states and transformations
taken together can be described in a unified manner using the
group $SO(8,{\mathbb C})$.

 Let us discuss the structure of four-qubit $SL(2,\mathbb
C)^{\times 4}$ invariants\cite{Luque,Levay4,Djokovic,Osterloh}.
The number of algebraically independent four-qubit invariants is
four. We have one quadratic, two quartic, and one sextic
invariant. In our recent paper\cite{Levay4} we investigated the
structure of these invariants in the special frame where two of
our qubits played a distinguished role. As we will see this
scenario is just the one needed in the two-center STU black hole
context since in this setting one of the special qubits (the one
labelled by the number $0$) will be associated to the horizontal
$SL_h(2,\mathbb R)$ of Ferrara et.al.\cite{Ferretal} and the
other (the one labelled by the number $1$) is arising as the first
factor from the structure $SL(2,\mathbb R)\times SO(2,2)$ known
from the STU model. Indeed such structure is the one arising as a
special case of the infinite Jordan symmetric sequence of $N=2$
$d=4$ supergravity theories\cite{seq}.

To an arbitrary state $\vert\Lambda\rangle$ we can also associate
a $4\times 4$ matrix \beq {\cal L}\equiv
\begin {pmatrix}{\Lambda}_{0000}&{\Lambda}_{0001}&{\Lambda}_{0010}&{\Lambda}_{0011}\\
{\Lambda}_{0100}&{\Lambda}_{0101}&{\Lambda}_{0110}&{\Lambda}_{0111}\\
{\Lambda}_{1000}&{\Lambda}_{1001}&{\Lambda}_{1010}&{\Lambda}_{1011}\\
{\Lambda}_{1100}&{\Lambda}_{1101}&{\Lambda}_{1110}&{\Lambda}_{1111}\end{pmatrix}
\equiv
\begin{pmatrix}A^1&A^2&A^3&A^4\\
B^1&B^2&B^3&B^4\\
C^1&C^2&C^3&C^4\\
D^1&D^2&D^3&D^4\end{pmatrix}, \label{fourvectors} \eeq \noindent
or four four-vectors. The splitting of the amplitudes of
$\vert\Lambda\rangle$ into such four-vectors reflects our special
choice for the distinguished qubits compatible with our
conventions. We will also need the matrices \beq
{\cal M}=\begin{pmatrix}A^1&A^2&B^1&B^2\\
C^1&C^2&D^1&D^2\\
A^3&A^4&B^3&B^4\\
C^3&C^4&D^3&D^4\end{pmatrix},\qquad {\cal N}=
\begin{pmatrix}A^1&A^3&B^1&B^3\\
A^2&A^4&B^2&B^4\\
C^1&C^3&D^1&D^3\\
C^2&C^4&D^2&D^4\end{pmatrix}.\label{matrices} \eeq \noindent
Notice that in four-qubit notation the matrices ${\cal M}$ and
${\cal N}$ are arising from ${\cal L}$ by the permutations
$(012)(3)$ and $(0)(123)$ meaning that the index structure of
these matrices is\beq {\cal L}\leftrightarrow
{\Lambda}_{i_0i_1i_2i_3},\qquad {\cal M}\leftrightarrow
{\Lambda}_{i_1i_2i_0i_3},\qquad\ {\cal N}\leftrightarrow
{\Lambda}_{i_0i_2i_3i_1}.\label{permutaciok}\eeq\noindent Notice
that these matrices are entering in the reduced density matrices
as \beq {\varrho}_{01}\equiv {\rm
Tr}_{23}\vert\Lambda\rangle\langle\Lambda\vert={\cal L}{\cal
L}^{\dagger}\qquad {\varrho}_{12}\equiv {\rm
Tr}_{03}\vert\Lambda\rangle\langle\Lambda\vert={\cal M}{\cal
M}^{\dagger}\qquad {\varrho}_{02}\equiv {\rm
Tr}_{13}\vert\Lambda\rangle\langle\Lambda\vert={\cal N}{\cal
N}^{\dagger}\label{redsur}\eeq\noindent \beq
\overline{{\varrho}}_{23}\equiv {\rm
Tr}_{01}\vert\Lambda\rangle\langle\Lambda\vert={\cal
L}^{\dagger}{\cal L}\qquad \overline{{\varrho}}_{03}\equiv {\rm
Tr}_{12}\vert\Lambda\rangle\langle\Lambda\vert={\cal
M}^{\dagger}{\cal M}\qquad \overline{{\varrho}}_{13}\equiv {\rm
Tr}_{02}\vert\Lambda\rangle\langle\Lambda\vert={\cal
N}^{\dagger}{\cal N}.\label{redsur2}\eeq\noindent where overline
denotes complex conjugation.

 Now we introduce on the vector space ${\mathbb
C}^4\simeq {\mathbb C}^2\times {\mathbb C}^2$ corresponding to the
third and fourth qubit a symmetric bilinear form ${\bf g}:{\mathbb
C}^4\times{\mathbb C}^4\to{\mathbb C}$ with matrix representation

\beq g={\varepsilon}\otimes
{\varepsilon}=\begin{pmatrix}0&1\\-1&0\end{pmatrix}\otimes
\begin{pmatrix}0&1\\-1&0\end{pmatrix}.\eeq\noindent
 This means that we
have an $SL(2,\mathbb C)^{\times 2}$ invariant quantity with the
explicit form \beq g(A,B)\equiv A\cdot B=
g_{\alpha\beta}A^{\alpha}B^{\beta}=A_{\alpha}B^{\alpha}=
A^1B^4-A^2B^3-A^3B^2+A^4B^1. \label{g} \eeq \noindent

We can also introduce a {\it dual four-qubit state} \beq
\vert\lambda\rangle=\sum_{i_0i_1i_2i_3=0,1}\lambda_{i_0i_1i_2i_3}\vert
i_0i_1i_2i_3\rangle \eeq \noindent with the associated matrix \beq
{\it l}\equiv
\begin {pmatrix}{\lambda}_{0000}&{\lambda}_{0001}&{\lambda}_{0010}&{\lambda}_{0011}\\
                {\lambda}_{0100}&{\lambda}_{0101}&{\lambda}_{0110}&{\lambda}_{0111}\\
           {\lambda}_{1000}&{\lambda}_{1001}&{\lambda}_{1010}&{\lambda}_{1011}\\
           {\lambda}_{1100}&{\lambda}_{1101}&{\lambda}_{1110}&{\lambda}_{1111}\end{pmatrix}
           \equiv\begin{pmatrix}a^1&a^2&a^3&a^4\\b^1&b^2&b^3&b^4\\
                                                  c^1&c^2&c^3&c^4\\
 d^1&d^2&d^3&d^4\end{pmatrix},\label{fourvectorsdual}\eeq\noindent
where \beq
a^{\alpha}={\epsilon}^{\alpha\beta\gamma\delta}B_{\beta}C_{\gamma}D_{\delta},\quad
b^{\beta}={\epsilon}^{\alpha\beta\gamma\delta}A_{\alpha}C_{\gamma}D_{\delta}\quad
c^{\gamma}={\epsilon}^{\alpha\beta\gamma\delta}A_{\alpha}B_{\beta}D_{\delta}\quad
d^{\delta}={\epsilon}^{\alpha\beta\gamma\delta}A_{\alpha}B_{\beta}C_{\gamma}.
\label{dual} \eeq \noindent Here  ${\epsilon}^{1234}=+1$, and
indices are lowered by the matrix of $g$. Notice that the
amplitudes of the dual four-qubit state are {\it cubic} in the
original ones.

Using these definitions we define the quadratic and sextic
invariants as \beq I_1\equiv \frac{1}{2}(A\cdot D-B\cdot C),\qquad
I_3\equiv \frac{1}{2}(a\cdot d-b\cdot c). \label{26} \eeq
\noindent The explicit form of the sextic invariant in terms of the
dot product of Eq.(\ref{g}) is \beq
2I_3={\rm Det}\begin{pmatrix}A\cdot A&A\cdot B&A\cdot D\\
A\cdot C&B\cdot C&C\cdot D\\
A\cdot D&B\cdot D&D\cdot D\end{pmatrix}-
{\rm Det}\begin{pmatrix}A\cdot B&B\cdot B&B\cdot C\\
A\cdot C&B\cdot C&C\cdot C\\
A\cdot D&B\cdot D&C\cdot D\end{pmatrix}. \label{sexkiirva} \eeq
\noindent We also recall that the explicit form of $I_1$ is hiding
its permutation invariance. Moreover, though the expression of
$I_3$ of Eq.(\ref{26}) is similar to the one of $I_1$ the
invariant $I_3$ is {\it not} invariant under the permutation of
the qubits.

Now we turn to the structure of quartic invariants. We have two
independent of such invariants\cite{Luque} and the simplest of
them is the obvious expression \beq I_4\equiv {\rm Det}{\cal L}
\label{detinv} \eeq \noindent i.e. the determinant of the $4\times
4$ matrix of Eq.(\ref{fourvectors}). In order to present the
definition of the second one we define separable bivectors of the
form \beq \Pi_{\mu\nu\alpha\beta}\equiv
\Lambda_{\mu\alpha}\Lambda_{\nu\beta}-\Lambda_{\mu\beta}\Lambda_{\nu\alpha},\qquad
\alpha,\beta,\mu,\nu=1,2,3,4. \label{bivectors1} \eeq \noindent
Here our labelling convention $\Lambda_{\mu\alpha}$ indicates that
$\mu=1,2,3,4$ identifies the four-vector in question (i.e. $A,B,C$
or $D$ of Eq.(\ref{fourvectors})), and the label $\alpha=1,2,3,4$
refers to the component of the particular vector. Now our last
invariant is the quartic combination \beq
I_2=\frac{1}{6}\Pi_{\mu\nu\alpha\beta}\Pi^{\mu\nu\alpha\beta}.
\label{last4} \eeq \noindent
Obviously the symmetric
nondegenerate bilinear form $g$ of Eq.(\ref{g}) acting on four
vectors like $A,B,C,D\in {\mathbb C}^4$ is inducing a
corresponding symmetric nondegenerate bilinear form on the space
of bivectors ${\bigwedge}^2{\mathbb C}^4$. By an abuse of notation
we use again the symbol $\cdot$ for this new bilinear form with
the definition\cite{Levay4} 
\beq (A\wedge B)\cdot (C\wedge D)\equiv
2((A\cdot C)(B\cdot C)-(A\cdot D)(B\cdot
C)).\label{wg}\eeq\noindent 
Now $I_2$ can also be written in the equivalent form 
\beq I_2=\frac{1}{6}\left[(A\wedge B)\cdot(C\wedge D)+(A\wedge
C)\cdot(B\wedge D)-\frac{1}{2}(A\wedge D)^2 -\frac{1}{2}(B\wedge
C)^2\right]. \label{explutolso} \eeq \noindent

An important comment here is in order. Let us have a look at $I_4$
and also at the determinants of the matrices of
Eq.(\ref{matrices}) \beq L\equiv I_4={\rm Det}{\cal L},\qquad
M\equiv {\rm Det}{\cal M},\qquad N\equiv{\rm Det}{\cal
N}.\label{detek}\eeq\noindent Then one can
prove\cite{Brylinski,Luque} \beq
L+M+N=0.\label{osszef}\eeq\noindent One also has the constraint
\beq M-N=3I_2-2I_1^2,\label{constraint}\eeq\noindent that we will
need later.

It is known\cite{Luque} that the minimal set of algebraically
independent $SL(2)^{\times 4}$ invariants is consisting of a
quadratic, two quartic and one sextic invariant. Our choice for
this set will be\cite{Levay4}: $I_1,I_2,I_4$ and $ I_3$. Let us
now present the reason for this choice. Let us consider the matrix
\beq \Omega\equiv {\cal L}g{\cal L}^T g. \label{Omega} \eeq
\noindent Then its characteristic polynomial is \beq
\Sigma_4(\Lambda_{i_0i_1i_2i_3},t)\equiv{\rm Det}({\bf
1}t-\Omega)=t^4-4I_1t^3+6I_2t^2-4I_3t+I_4^2. \label{poli1} \eeq
\noindent Clearly by Newton's identities we have \beq
I_1=\frac{1}{4}{\rm Tr}\Omega,\qquad I_2=\frac{1}{12}[({\rm
Tr}\Omega)^2-{\rm Tr}\Omega^2], \eeq \noindent \beq
I_3=\frac{1}{24}[({\rm Tr}\Omega)^3-3{\rm Tr}\Omega{\rm
Tr}\Omega^2+2{\rm Tr}\Omega^3],\qquad (I_4)^2={\rm Det}\Omega.
\eeq \noindent This form of writing our invariants is related to
the fact that there is a $1-1$ correspondence between the
$SL(2,{\mathbb C})^{\otimes 4}$ orbits of four-qubit states and
the $SO(4,\mathbb C)\times SO(4,\mathbb C)$ ones of $4\times 4$
matrices.

The polynomial of Eq.(\ref{poli1}) in the four-qubit context
appeared in our recent paper\cite{Levay4} its role as a
characteristic polynomial has been emphasized in
Ref.\cite{Djokovic}. The discriminant of this fourth order
polynomial is the hyperdeterminant\cite{Zelevinsky} $D_4$ of the
$2\times 2\times 2\times 2$ hypercube $\Lambda_{i_0i_1i_2i_3}$. It
is a polynomial of degree $24$ in the $16$ amplitudes and has
2894276 terms\cite{Debbie}. $D_4$ can be expressed\cite{Levay4} in
terms of our fundamental invariants as \beq 256D_4=S^3-27T^2
\label{hyper4} \eeq \noindent where \beq
S=(I_4^2-I_2^2)+4(I_2^2-I_1I_3),\quad
T=(I_4^2-I_2^2)(I_1^2-I_2)+(I_3-I_1I_2)^2.\label{ST} \eeq
\noindent For an alternative form of $D_4$ see the papers
 of Refs.\cite{Luque,Osterloh}.

 In closing this section we briefly discuss
some results on the classification of entanglement classes for
four qubits\cite{Verstraete,Djokovic}. By entanglement classes we
mean orbits under $SL(2,{\mathbb C})^{\times 4}\cdot {\rm Sym}_4$
where ${\rm Sym}_4$ is the symmetric group on four symbols. The
basic result states that four qubits can be entangled in nine
different ways\cite{Verstraete,Djokovic}. It is to be contrasted
with the two entanglement classes\cite{Dur} obtained for three
qubits. For a refined classification of four qubit entanglement
motivated by the black hole qubit correspondence see the papers of
Borsten et.al.\cite{Borstenfour}

Let us consider the matrix \beq {\cal
R}_{\Lambda}\equiv\begin{pmatrix}0&\Lambda
g\\-\Lambda^Tg&0\end{pmatrix} \label{rmatrix} \eeq \noindent 
which now can be regarded as an element of the Lie algebra of $SO(8,{\mathbb C})$. If
the matrix ${\cal R}_{\Lambda}$ is diagonalizable under the action
\beq {\cal R}_{\Lambda}\mapsto S{\cal R}_{\Lambda}S^{-1},\qquad
S=\begin{pmatrix}S_0\otimes S_1&0\\0&S_2\otimes
S_3\end{pmatrix},\qquad S_{\alpha}\in SL(2,{\mathbb C}) \eeq
\noindent we say that the corresponding four-qubit state
$\vert\Lambda\rangle$ is {\it semisimple}. If ${\cal R}_{\Lambda}$
is {\it nilpotent} then we call the corresponding state
$\vert\Lambda\rangle$ nilpotent too. It is known that a nilpotent
orbit is {\it conical} i.e. if $\vert\Lambda\rangle$ is an element
of the orbit then $\lambda\vert\Lambda\rangle$ is also an element for
all nonzero complex numbers $\lambda$. Hence a  nilpotent orbit is also
a $GL(2,{\mathbb C})^{\times 4}$ orbit. It is clear that for
nilpotent states all of our algebraically independent invariants
are zero.

A semisimple state of four qubits can always be
transformed to the form\cite{Verstraete}
\begin{eqnarray}
\vert G_{abcd}\rangle&=&\frac{a+d}{2}(\vert 0000\rangle +\vert
1111\rangle) +\frac{a-d}{2}(\vert 0011\rangle +\vert
1100\rangle)\nonumber\\&+& \frac{b+c}{2}(\vert 0101\rangle +\vert
1010\rangle)+ \frac{b-c}{2}(\vert 0110\rangle +\vert 1001\rangle),
\label{genuine}
\end{eqnarray}
\noindent where $a,b,c,d$ are complex numbers. This class
corresponds to the so called GHZ class found in the three-qubit
case\cite{Dur}. For this state the reduced density matrices
obtained by tracing out all but one of the qubits are proportional
to the identity. This is the state with maximal four-partite
entanglement. Another interesting property of this state is that
it does not contain true three-partite entanglement. A
straightforward calculation shows that the values of our
invariants $(I_1,I_2,I_3,I_4)$ occurring for the state $\vert
G_{abcd}\rangle$ representing the generic class are \beq
I_1=\frac{1}{4}[a^2+b^2+c^2+d^2],\quad
I_2=\frac{1}{6}[(ab)^2+(ac)^2+(ad)^2+(bc)^2+(bd)^2+(cd)^2], \eeq
\noindent \beq
I_3=\frac{1}{4}[(abc)^2+(abd)^2+(acd)^2+(bcd)^2],\quad I_4=abcd
\label{genpar}\eeq \noindent hence the values of the invariants
$(4I_1,6I_2,4I_3,I_4^2)$ are given in terms of the elementary
symmetric polynomials in the variables
$(t_1,t_2,t_3,t_4)=(a^2,b^2,c^2,d^2)$. For the semisimple states $\vert
G_{abcd}\rangle$ the value of $D_4$ can be expressed
as\cite{Luque,Levay4} \beq D_4=\frac{1}{256}\Pi_{i<j}(t_i-t_j)^2,
\qquad (t_1,t_2,t_3,t_4)\equiv
(a^2,b^2,c^2,d^2).\label{gyokok}\eeq\noindent
Notice that for the states $\vert G_{abcd}\rangle$ 
with $D_4$ nonvanishing ($t_i\neq t_j$) the corresponding matrix of Eq. (\ref{rmatrix}) belongs to a Cartan subalgebra of $SO(8,{\mathbb C})$.
The stabilizer of such states corresponds to the Weyl group of $SO(8,{\mathbb C})$.
This stabilizer is the Klein group generated by the four elements
$I\otimes I\otimes I\otimes I$ and $\sigma_a\otimes \sigma_a\otimes \sigma_a\otimes \sigma_a$ for $a=1,2,3$.

\section{Two center extremal black holes as four qubits systems.}

In the paper\cite{Ferretal} of Ferrara et.al. in order to describe the structure
of the U-duality invariant polynomials associated to the
two-center extremal black hole solutions the Calabi-Visentini (CV)
basis has been used.
Here by $U$-duality we mean the continuous limit\cite{Ferretal}
valid for large values of the charges of the usual-nonperturbative
string theory symmetries.
First we describe the connection of the CV
basis to the one making the four-qubit structures explicit. Next
we turn to an entanglement based understanding of the structure of
the two-center $U$-duality invariants.

In the CV basis the two-center black hole solutions are
characterized by a pair of {\it real} charge vectors ${\cal Q}_1,
{\cal Q}_2\in{\mathbb R}^8$

\beq {\cal Q}_1\equiv (P^0,P^1,P^2,P^3,Q_0,Q_1,Q_2,Q_3)^T,\quad
{\cal Q}_2\equiv (p^0,p^1,p^2,p^3,q_0,q_1,q_2,q_3)^T.
\label{charge}\eeq \noindent

 As we see these charge vectors are containing two
four-vectors each namely $P^I$, $Q_I\equiv {\eta}_{IJ}Q^J$, and
$p^I, q_I\equiv {\eta}_{IJ}q^J$ $I,J=0,1,2,3$ where the raising
and lowering of the indices $I$ and $J$ are effected by the metric
${\eta}_{IJ}$ and ${\eta}^{IJ}$ of $SO(2,2)$ answering the
symmetric bilinear form  $h$ acting on the charge four-vectors
$P^{I}$ and $Q^{J}$  in the CV basis as \beq h(P,Q)\equiv P\circ
Q={\eta}_{IJ}P^IQ^J=-P^0Q^0-P^1Q^1+P^2Q^2+P^3Q^3. \eeq \noindent

Let us now relate the $16$ component charge vector in the CV basis
characterizing a particular two center extremal black hole
solution in the STU model to a {\it real unnormalized} four-qubit
pure state $\vert \Lambda\rangle $ by relating the four-vectors
$P^I,Q^I,p^I,q^I\in {\mathbb R}^4$ of Eq.(\ref{charge}) and
$A^{\alpha},B^{\alpha},C^{\alpha},D^{\alpha}\in{\mathbb R}^4$ of
Eq.(\ref{fourvectors}) as follows.

\beq\begin{pmatrix}P^0&\\P^1\\P^2\\P^3\end{pmatrix}_{CV}=\frac{1}{\sqrt{2}}
\begin{pmatrix}A^1-A^4\\A^2+A^3\\-A^1-A^4\\-A^2+A^3\end{pmatrix},\qquad
\begin{pmatrix}Q^0&\\Q^1\\Q^2\\Q^3\end{pmatrix}_{CV}=\frac{1}{\sqrt{2}}
\begin{pmatrix}B^1-B^4\\B^2+B^3\\-B^1-B^4\\-B^2+B^3\end{pmatrix}\label{atteres1}
\eeq \noindent

\beq\begin{pmatrix}p^0&\\p^1\\p^2\\p^3\end{pmatrix}_{CV}=\frac{1}{\sqrt{2}}
\begin{pmatrix}C^1-C^4\\C^2+C^3\\-C^1-C^4\\-C^2+C^3\end{pmatrix},\qquad
\begin{pmatrix}q^0&\\q^1\\q^2\\q^3\end{pmatrix}_{CV}=\frac{1}{\sqrt{2}}
\begin{pmatrix}D^1-D^4\\D^2+D^3\\-D^1-D^4\\-D^2+D^3\end{pmatrix}
\label{atteres2}\eeq \noindent Then using the definitions above we
clearly have for example \beq P\circ Q=A\cdot
B=g(A,B)=g_{\alpha\beta}A^{\alpha}B^{\beta} \eeq\noindent with the
bilinear form $g$ defined as in Eq.(\ref{g}).

Let us also give the connection between the Calabi-Visentini basis
and the one usually used in special geometry, i.e. the special
coordinates (SC) symplectic frame. This frame yields the usual set
of electric and magnetic charges i.e. $(P^I,Q_I)_{SC}$ and
$(p^I,q_I)_{SC}$. In the following we will use these charges so it
is important to clarify their relationship to the components of
our four qubit state $\vert\Lambda\rangle$ (see
Eq.(\ref{fourvectors})). \beq
\begin{pmatrix}\Lambda_{0000}\\\Lambda_{0001}\\\Lambda_{0010}\\\Lambda_{0011}\end{pmatrix}=
\begin{pmatrix}A^1\\A^2\\A^3\\A^4\end{pmatrix} =\begin{pmatrix}
P^0\\P^2\\P^3\\Q_1\end{pmatrix}_{SC},\qquad
\begin{pmatrix}\Lambda_{0100}\\\Lambda_{0101}\\\Lambda_{0110}\\\Lambda_{0111}\end{pmatrix}=
\begin{pmatrix}B^1\\B^2\\B^3\\B^4\end{pmatrix} =\begin{pmatrix}
P^1\\Q_3\\Q_2\\-Q_0\end{pmatrix}_{SC} \label{atteres3} \eeq\
\noindent

\beq
\begin{pmatrix}\Lambda_{1000}\\\Lambda_{1001}\\\Lambda_{1010}\\\Lambda_{1011}\end{pmatrix}=
\begin{pmatrix}C^1\\C^2\\C^3\\C^4\end{pmatrix} =\begin{pmatrix}
p^0\\p^2\\p^3\\q_1\end{pmatrix}_{SC},\qquad
\begin{pmatrix}\Lambda_{1100}\\\Lambda_{1101}\\\Lambda_{1110}\\\Lambda_{1111}\end{pmatrix}=
\begin{pmatrix}D^1\\D^2\\D^3\\D^4\end{pmatrix} =\begin{pmatrix}
p^1\\q_3\\q_2\\-q_0\end{pmatrix}_{SC}. \label{atteres4} \eeq\
\noindent We note here however, that our conventions are slightly
different from the ones used in Ref.1. The charges
$p^1,p^2,p^3,q_0$ and $P^1,P^2,P^3,Q_0$ in the SC basis used by us
are the negatives of the corresponding ones in the SC basis as
used in Ref.1. see Eqs.(\ref{atteres1}-\ref{atteres2}),
Eqs.(\ref{atteres3}-\ref{atteres4}) and Eq. (B.3) of that paper.

Notice also that in our four-qubit state $\vert\Lambda\rangle$
with amplitudes $\Lambda_{i_0i_1i_2i_3}$ {sit {\it two}
three-qubit states with amplitudes $\Lambda_{0i_1i_2i_3}$, and
$\Lambda_{1i_1i_2i_3}$. The first set of amplitudes labelled by
$i_0=0$ corresponds to the charge configuration of the first black
hole and the second labelled by $i_0=1$ describes the second black
hole. As we see the first label plays a distinguished role with an
extra $SL(2,{\mathbb R})_0$ (dubbed by Ferrara et.al. the
horizontal one) acting on. This group represents the generalized exchange
symmetry between the centers.

Let us now define three  bivectors \beq X\equiv A\wedge B,\qquad
Y\equiv C\wedge D,\qquad Z\equiv \frac{1}{2}(A\wedge D-B\wedge C).
\label{bivectors} \eeq \noindent In component notation we have for
example \beq X_{\alpha\beta
}=A_{\alpha}B_{\beta}-A_{\beta}B_{\alpha}.\eeq\noindent Switching
to the Calabi-Visentini basis these objects are the $T$-tensors
used in the paper of Ferrara et.al.\cite{Ferretal} As we can see
the bivectors $X$ and $Y$ are separable i.e. they are precisely
the ones satisfying the Pl\"ucker relations \beq
X_{12}X_{34}-X_{13}X_{23}+X_{14}X_{23}=0,\qquad
Y_{12}Y_{34}-Y_{13}Y_{23}+Y_{14}Y_{23}=0
\label{Plucker}\eeq\noindent on the other hand the bivector $Z$ is
{\it entangled} i.e. in the nomenclature of fermionic
entanglement\cite{Pipek} it has Slater rank two.

Now we introduce the shorthand notation for the product of two
separable bivectors as defined in Eq.(\ref{wg}) \beq X\cdot
Y\equiv (A\wedge B)\cdot(C\wedge D)\label{wg1}\eeq\noindent Notice
that in the notation of Eq.(\ref{atteres3})-(\ref{atteres4}) \beq
X^2\equiv X\cdot X=2(A^2B^2-(A\cdot B)^2),\qquad Y^2\equiv Y\cdot
Y=2(C^2D^2-(C\cdot D)^2) \eeq\noindent are just two times the
quartic invariants of the charges characterizing the two black
holes \beq \frac{1}{2}X^2=I_4({\cal
Q}_1)=-D_3(\Lambda_{0i_1i_2i_3}),\qquad \frac{1}{2}Y^2=I_4({\cal
Q}_2)=-D_3(\Lambda_{1i_1i_2i_3}).\eeq \noindent Here
$D_3(\Lambda_{i_1i_2i_3})$ is Cayley's
hyperdeterminant\cite{Cayley}.

With these definitions we can define the quantities \beq
I_{+2}=\frac{1}{2}X^2,\quad I_{+1}=\frac{1}{2}X\cdot Z,\quad
I_{0}=\frac{1}{6}(2Z^2-X\cdot Y),\quad I_{-1}=\frac{1}{2}Y\cdot
Z,\quad I_{-2}=\frac{1}{2}Y^2\label{inv1}\eeq\noindent and the
ones \beq I^{\prime}=\frac{1}{2}X\cdot Y,\qquad
I^{\prime\prime}=\frac{3}{2}Z^2.\label{inv2}\eeq\noindent When
reinterpreted in the CV basis these are precisely the
$SL(2,{\mathbb R})_1\times SL(2,{\mathbb R})_2\times SL(2,{\mathbb
R})_3$ invariants of Ferrara et.al.\cite{Ferretal} The important property of
the invariants of Eq.(\ref{inv1}) is that they are {\it
covariants} with respect to $SL(2,{\mathbb R})_0$ acting on the
distinguished (horizontal) qubit.
Indeed, they are sitting in the ${\bf 5}$ (spin $2$) irreducible
representation of this group.

Now we elucidate another aspect of this important property of the
set of invariants of Eq.(\ref{inv1}). In order to do this we first
recall that for two-qubits the canonical measure of pure state
entanglement is the concurrence\cite{Kundu} \beq {\cal C}=2\vert
D_2(\Lambda_{i_1i_2})\vert=2\vert
\Lambda_{00}\Lambda_{11}-\Lambda_{01}\Lambda_{10}\vert\label{conc}\eeq\noindent
which is related to the determinant of an ordinary $2\times 2$
matrix. For three-qubits the basic quantity characterizing genuine
three-qubit entanglement\cite{Dur} is the three-tangle \beq
\tau=4\vert D_3(\Lambda_{i_0i_1i_2})\vert\eeq\noindent where now
$D_3$ is Cayley's hyperdeterminant\cite{Cayley,Kundu}. According
to the method of Schl\"afli\cite{Zelevinsky} $D_3$ is related to
the discriminant $\Delta_2$ of the quadratic polynomial \beq
\Pi_2( \Lambda_{i_0i_1i_2},t)\equiv
D_2(\Lambda_{0i_1i_2}t+\Lambda_{1i_1i_2})=(\Lambda_0\cdot\Lambda_0)t^2+2(\Lambda_0\cdot\Lambda_1)t+
(\Lambda_1\cdot\Lambda_1)\eeq\noindent where $\Lambda_0$ and
$\Lambda_1$ are four-vectors with components
$(\Lambda_{000},\Lambda_{001},\Lambda_{010},\Lambda_{011})$ and
$(\Lambda_{100},\Lambda_{101},\Lambda_{110},\Lambda_{111})$ and
the $\cdot$ product is the usual one of Eq.(\ref{g}). Moreover,
due to permutation invariance of $D_3$ we obtain the same
expression whenever the first or the second qubit plays a
distinguished role. Obviously the quantities
$J_{+1}=(\Lambda_0\cdot\Lambda_0)$,
$J_0=(\Lambda_0\cdot\Lambda_1)$, and
$J_{-1}=(\Lambda_1\cdot\Lambda_1)$ are $SL(2)_1\times SL(2)_2$
invariants, however they are {\it covariants} under the
"horizontal" $SL(2)_0$. Indeed, the triple $J_{+1},J_0,J_{-1}$
transforms according to the irreducible representation ${\bf 3}$
of spin $1$ of this "horizontal" group which can be regarded as
some sort of generalized exchange symmetry working between the
{\it two} two-qubit systems. It is also clear that searching
singlets with respect to this "horizontal" symmetry group can
reveal some new properties of our pair of two qubit systems,
namely that they are secretly comprising a system having a higher
degree of symmetry.

Now proceeding by analogy we define the polynomial \beq
\Pi_4(\Lambda_{i_0i_1i_2i_3},t)\equiv
D_3(\Lambda_{0i_1i_2i_3}t+\Lambda_{1i_1i_2i_3})=I_{+2}t^4+4I_{+1}t^3+6I_0t^2+4I_{-1}t+I_{-2}.
\label{poli2}\eeq\noindent A straightforward calculation shows
that the coefficients of this polynomial are precisely the
covariants of Eq.(\ref{inv1}). Now according to theorem 14.4.1 and
corollary 14.2.10 of Ref.\cite{Zelevinsky} the discriminant of
this quartic polynomial $\Delta_4$ divided by $256$ is just the
hyperdeterminant $D_4(\Lambda_{i_0i_1i_2i_3})$ of the hypercube of
format $2\times 2\times 2\times 2$. Besides being a singlet with
respect to the horizontal $SL(2)_0$, $D_4$ is also an invariant
under $S_4$ the permutation group of the four qubit system.

At this point one can notice\cite{Levay4} that the polynomials
$\Sigma_4$ and $\Pi_4$ of Eqs.(\ref{poli1}) and (\ref{poli2}) are
having the same discriminants $\Delta_4$ hence both can be used to
obtain an expression for $D_4$. Notice that $\Sigma_4$ is a
polynomial with its coefficients $I_1,I_2,I_3,I_4$ also being
$SL(2)_0$ singlets however, $\Pi_4$ is a polynomial with
coefficients $I_{\pm 2},I_{\pm 1}, I_0$ being merely $SL(2)_0$
covariants. Using this observation we can construct new $SL(2)_0$
singlets from the quantities of Eq.(\ref{inv1}) by relating them
to the known algebraically independent four-qubit ones namely
$I_1,I_2,I_3$ and $I_4$.

In order to relate the $SL(2)_0$ singlets found by Ferrara et.al.
to our four algebraically independent four-qubit invariants one
just has to compare the relevant expressions. In fact many of
their invariants and the constraints satisfied by them take in
this four-qubit setting a much simpler and instructive form. In
particular their complete set of invariants with corresponding
degrees $2,4,6,8$ denoted by ${\cal W},\chi, I_6,{\rm Tr}({\cal
J}^2)$ is related to ours as \beq {\cal W}=2I_1,\qquad
\chi=3I_2-2I_1^2,\qquad I_6=-I_3.\label{dictionary}\eeq\noindent
By virtue of Eqs. (\ref{constraint}) and (\ref{inv2}) we also have
the relations \beq
I^{\prime}-I^{\prime\prime}=\frac{3}{2}I_2,\qquad
\chi=M-N.\label{kellenifog}\eeq\noindent

We still have to account for the invariant ${\rm Tr}({\cal J}^2)$
of order $8$ built from ${\cal J}$ the symmetric traceless matrix
comprising the covariants $I_{\pm 2},I_{\pm 1},I_0$. The $5$
independent components of this matrix are transforming according
to the ${\bf 5}$ of $SL(2)_0$. In order to reveal the meaning of
this invariant and also an extra one ${\rm Tr}({\cal J}^3)$ of
order $12$ let us reconsider the awkward looking polynomial
constraint of Eq. (5.6) of Ref.1.

\beq {\cal P}_{12}\equiv I_6^2+{\cal W}{\chi}I_6+{\rm Tr}({\cal
J}^3)+ \frac{{\rm Tr}({\cal J}^2){\cal W}^2}{12} -\frac{{\rm
Tr}({\cal J}^2)\chi}{3}-\frac{{\cal W}^6}{432}+\frac{{\cal
W}^4\chi}{36}+\frac{5{\cal
W}^2{\chi}^2}{36}+\frac{4{\chi}^3}{27}=0. \label{p12}\eeq\noindent Using the
dictionary of Eq. (\ref{dictionary}) we can cast this constraint
in the nice form \beq {\rm Tr}({\cal J}^3)=[4(I_2^2-I_1I_3)-{\rm
Tr}({\cal J}^2)](I_1^2-I_2)-(I_3-I_1I_2)^2.\label{jolvan1}
\eeq\noindent

There is one more invariant\cite{Ferretal} of order $8$ which is
directly related to ${\rm Tr}({\cal J}^2)$ \beq {\cal P}_8\equiv
-12{\rm Tr}({\cal J}^2)+24I_6{\cal W}+({\cal
W}^2+2\chi)^2.\eeq\noindent Recall that the constraint ${\cal
P}_8=0$ implements the reduction of the $stu$ model to the $st^2$
model\cite{Ferretal} in a manifestly $SL(2)_0$ invariant manner.
Using again Eq.(\ref{dictionary}) the new form of ${\cal P}_8$ is
\beq {\cal P}_8=12(3I_2^2-4I_1I_3-{\rm Tr}({\cal
J}^2)).\label{jolvan2}\eeq\noindent Putting this into
Eq.(\ref{jolvan1}) and recalling
 Eq.(\ref{ST}) one obtains the simple expressions \beq {\rm Tr}({\cal J}^2)=S,\qquad {\rm Tr}({\cal
J}^3)=-T,\label{jolvan3}\eeq\noindent provided \beq {\cal
P}_8=-{12}I_4^2=-12L^2.\label{jolvan4}\eeq\noindent

The first result of these considerations is that
$-\frac{1}{12}{\cal P}_8$ is really the square of the basic fourth
order invariant $L=I_4$. According to Eqs.(\ref{fourvectors}) and
(\ref{detek}) $L$ is just the determinant of the matrix ${\cal L}$
we have started our four-qubit considerations with. Moreover,
according to Eqs.(\ref{redsur}-\ref{redsur2}) we also see that for
{\it unnormalized} four-qubit states we have \beq
-\frac{1}{12}{\cal P}_8={\rm Det}{\varrho}_{01}={\rm
Det}{\varrho}_{12},\eeq\noindent hence this invariant is related
to the determinant of the reduced density matrices of our
four-qubit state $\vert\Lambda\rangle$ corresponding to the
bipartite split of the form: $(01)(23)$. We can also conclude that
in the four-qubit picture the reduction of the $stu$ model to the
$st^2$ is effected by sending one of the eigenvalues of the
reduced density matrices corresponding to the $(01)(23)$ split to
zero. Notice however, that the remaining density matrices
corresponding to the remaining two splits $(02)(13)$ and
$(03)(12)$ are generally not sharing this property.  This means
that for the $st^2$ model we have \beq L=0,\qquad M\neq 0,\qquad
N\neq 0.\eeq\noindent

Recall now that a suitable further reduction to the $t^3$ model is
obtained by employing the following two $SL(2)_0$ invariant
constraints\cite{Ferretal} \beq \chi=0,\qquad {\cal
P}_8=0.\eeq\noindent By virtue of Eq.(\ref{kellenifog}) and the
constraint $L+M+N=0$ these constraints can be described in the
compact form \beq L=M=N=0.\eeq \noindent This means that for the
$t^3$ reduction of the $stu$ model {\it all of the} reduced
density matrices of Eqs.(\ref{redsur}-\ref{redsur2}) of the four
qubit state $\vert\Lambda\rangle$ have a zero eigenvalue.

The second result of our considerations finally clarifies the role
played by the invariants ${\rm Tr}({\cal J}^2)$ and ${\rm
Tr}({\cal J}^3)$. In particular looking at Eqs.(\ref{hyper4}),
(\ref{ST}) and Eq.(\ref{jolvan3}) we see that that the four-qubit
hyperdeterminant $256D_4$ which is just the discriminant of our
polynomial $\Sigma_4$ of Eq.(\ref{poli1}) can be expressed with
the help of these invariants of order $8$ and $12$ as $S^3-27T^2$.
Moreover, since the discriminant of a quartic is the same as the
discriminant of its resolvent cubic one can also show that the two
quartic equations $\Sigma_4=0$ and $\Pi_4=0$ of Eqs. (\ref{poli1})
and (\ref{poli2}) are having the {\it same} resolvent cubics.
Indeed, a straightforward calculation shows that the corresponding
resolvent qubics in both cases are of the form \beq
u^3-Su-2T=0.\label{rcubic}\eeq\noindent In the first case we get
back to the known expressions for $S$ and $T$ of Eq.(\ref{ST}),
and in the second one we get\cite{Ferretal} \beq
S=3I_0^2-4I_{+1}I_{-1}+I_{+2}I_{-2},\qquad
T=I_0^3+I_{+1}^2I_{-2}+I_{-1}^2I_{+2}-I_{+2}I_0I_{-2}-2I_{+1}I_{0}I_{-1}.
\label{Suj}\eeq\noindent

\section{The $D2D6-D0D4$ split}
\subsection{Invariants}
In order to uncover the role of our four-qubit invariants playing
in the theory of two center black hole solutions let us consider a
special class of two center black hole solutions in the $t^3$
model featuring a $D0D4-D2D6$ split in the type $IIA$ duality frame. In this case the vectors of
Eq. (\ref{atteres3})-(\ref{atteres4}) in the special coordinate
(SC) basis are \beq
\begin{pmatrix}A^1\\A^2\\A^3\\A^4\end{pmatrix}=\begin{pmatrix}0\\P\\P\\0\end{pmatrix},\qquad
\begin{pmatrix}B^1\\B^2\\B^3\\B^4\end{pmatrix}=\begin{pmatrix}P\\0\\0\\-U\end{pmatrix}
\eeq\noindent corresponding to the first black hole with charge
configuration ${\cal Q}_1$ of a BPS $D0D4$ system and\beq
\begin{pmatrix}C^1\\C^2\\C^3\\C^4\end{pmatrix}=\begin{pmatrix}v\\0\\0\\q\end{pmatrix},\qquad
\begin{pmatrix}D^1\\D^2\\D^3\\D^4\end{pmatrix}=\begin{pmatrix}0\\q\\q\\0\end{pmatrix}
\eeq\noindent corresponding to the second black hole with charge
configuration ${\cal Q}_2$ of a BPS $D2D6$ system. For BPS
configurations in the first case we should have \beq -D_3({\cal
Q}_1)=(A\cdot A)(B\cdot B)-(A\cdot
B)^2=4UP^3>0\label{BPS1}\eeq\noindent and in the second the
corresponding constraint is \beq -D_3({\cal Q}_2)=(C\cdot
C)(D\cdot D)-(C\cdot D)^2=-4vq^3>0.\label{BPS2}\eeq\noindent
Notice that our charge split for the special values of $U=4$,
$P=q=1$ and $v=-4$ incorporates the illustrative example of Bates
and Denef\cite{Bates} (in that paper $v$ is related to ours via a
sign flip). The four-vectors $A^{\alpha},B^{\beta},C^{\gamma}$ and
$D^{\delta}$ are comprising the $16$ amplitudes of a real four-qubit
state as displayed in Eqs. (\ref{Lambda}) and (\ref{fourvectors}).
Now using Eqs. (\ref{26})-(\ref{sexkiirva}), (\ref{detinv}) and
(\ref{explutolso}) the algebraically independent four-qubit
invariants $I_1,I_2,I_3$ and $I_4$ can be calculated. The explicit
forms of these invariants are \beq I_1=\frac{1}{2}(Uv-3Pq),\qquad
I_2=\frac{2}{3}I_1^2,\qquad I_3=-Pq(Pq+Uv)^2,\qquad
I_4=0.\label{explinvar}\eeq\noindent

Notice that by virtue of Eqs.(\ref{osszef})-(\ref{constraint}) our
$D0D4-D2D6$ example illustrates the constraints we have already
discussed in connection with the $t^3$ model, namely the ones
$L=M=N=0$. Since $L=I_4=0$ and $3I_2=2I_1^2$ ($\chi=0$) the
quartic equation ${\Sigma}_4=0$ arising from the polynomial of Eq.
(\ref{poli1}) now reduces to a cubic one of the following form \beq
t^3-4I_1t^2+4I_1^2t-4I_3=0\label{poly3}\eeq\noindent The
discriminant of this cubic equation is \beq
\Delta=4I_3\left(I_3-\left(\frac{2I_1}{3}\right)^3\right).\label{cubicdiscr}\eeq\noindent
According to Eqs. (\ref{hyper4})-(\ref{ST}) the hyperdeterminant
$D_4$ is also related to this discriminant and is of the form \beq
256D_4=27I_3^3\left(\left(\frac{2I_1}{3}\right)^3-I_3\right)=
-(UvPq)(Pq)^2(Uv+9Pq)^2(Uv+Pq)^6.\label{explhyp}\eeq\noindent

\subsection{Consistency condition}
Let us now consider the necessary condition\cite{Bates,Denef2} for
our two center charge configuration supporting a corresponding two
center stationary extremal BPS black hole solution. As it is
well-known this condition is of the form \beq \vert{\bf x}_1-{\bf
x}_2\vert=\langle{\cal Q}_1,{\cal Q}_2\rangle\frac{\vert {\cal
Z}_1+{\cal Z}_2\vert_{r=\infty}}{2{\rm Im}(\overline{{\cal
Z}_2}{\cal Z}_1)_{r=\infty}}\label{necessary}\eeq\noindent where
${\bf x}_{1,2}$ are the locations of the centers, $r=\vert {\bf
x}\vert$, ${\cal Z}_{1,2}$ are the central charges corresponding
to the charges ${\cal Q}_{1,2}$, and the symplectic product of the
charge vectors $\langle{\cal Q}_1,{\cal Q}_2\rangle$ is related to
our quadratic four-qubit invariant as \beq
I_1=\frac{1}{2}\langle{\cal Q}_1,{\cal Q}_2\rangle.\eeq\noindent
The explicit forms of the central charges for our centers are \beq
{\cal Z}_1=e^{K/2}(U-3P{\tau}^2),\qquad {\cal
Z}_2=e^{K/2}(3q\tau+v{\tau}^3).\eeq\noindent Here $\tau$ is as
usual the complex scalar field of the $t^3$ model \beq
\tau=x-iy,\qquad y>0\eeq\noindent with negative imaginary
part\cite{Gimon,stu}, and $K=-\log(8y^3)$ is the K\"ahler
potential.

Now the constraint dictated by Eq.(\ref{necessary}) is \beq
I_1{\rm Im}(\overline{{\cal Z}}_2{\cal
Z}_1)_{\infty}>0.\label{nec2}\eeq\noindent Explicitly we have \beq
8y_{\infty}^2{\rm Im}(\overline{\cal Z}_2 {\cal
Z}_1)_{\infty}=-3Pv(x_{\infty}^2+y_{\infty}^2)^2+(9Pq+3Uv)x_{\infty}^2+
(9Pq-Uv)y_{\infty}^2+3Uq. \eeq\noindent We are
interested in the structure of BPS $D0D4-D2D6$ composites hence
according to Eqs.(\ref{BPS1})-(\ref{BPS2}) we should have \beq
UP>0,\qquad vq<0.\label{bps}\eeq\noindent For all possible sign
combinations satisfying these constraints we have \beq
(-Pv)I_1<0.\label{ezfontos}\eeq\noindent Let us now write our
consistency condition as \beq
8y_{\infty}^2\left(-\frac{I_1}{Pv}\right)\left(-Pv{\rm
Im}(\overline{Z}_2Z_1)_{\infty}\right)\equiv
\left(-\frac{I_1}{Pv}\right){\cal P}
>0.\eeq\noindent
Now a calculation shows that by virtue of Eq.(\ref{ezfontos})
consistency demands that \beq {\cal
P}=3\left(Pv(x_{\infty}^2+y_{\infty}^2)+I_1-Uv\right)^2+
UP(2vy_{\infty})^2-\frac{3}{4}(Uv+Pq)(Uv+9Pq)<0.\label{baromifontos}
\eeq\noindent

Let us now look at the expression for our hyperdeterminant of
format $2\times 2\times 2\times 2$ as given by Eq.(\ref{explhyp}).
Clearly positivity of $D_4$ implies that $-UvPq>0$ and neither
$Uv= -Pq$ nor $Uv=-9Pq$. Notice that all of these conditions are
compatible with our physically interesting situation. For
$-UvPq>0$ is compatible with our choice of signs supporting a pair
of BPS configurations, and in order to talk about a $D0D4-D2D6$
split none of the four charges can be zero. Moreover, since the
first term of Eq.(\ref{baromifontos}) is positive, the second one
for BPS configurations is positive as well ($UP>0$), in order to
satisfy this condition $(Uv+Pq)(Uv+9Pq)$ has to be positive. Hence we
see that {\it all} of the physically relevant conditions are
encoded into the structure of the hyperdeterminant $D_4$. In particular
$D_4>0$ and $-PvI_1<0$ of Eq.(\ref{ezfontos}) gives a
necessary condition for the consistency condition to hold.
Unfortunately the second of our conditions is featuring $-Pv$
which is not coming from any of our four-qubit invariants.

In order to eliminate this shortcoming let us now take another look
at the form of our polynomial of Eq.(\ref{poli1}). By employing
the substitution \beq t=I_1+y\eeq\noindent this polynomial can be
transformed to the reduced form \beq y^4+ay^2+by+c \eeq \noindent
where \beq a=6(I_2-I_1^2),\qquad b=12I_1I_2-4I_3-8I_1^3,\qquad
c=-3(I_1^2-I_2)^2+S \eeq\noindent and $S$ is our well-known
quantity defined by Eq.(\ref{ST}) or alternatively by
Eq.(\ref{Suj}). Now the conditions \beq \Delta_4>0,\qquad
a<0,\qquad c<\frac{a^2}{4}\eeq\noindent imply\cite{quart} that our
original polynomial Eq.(\ref{poli1}) featuring the fundamental
$4$-qubit invariants is having only {\it real} roots. Note that here
$\Delta_4$ is the discriminant of Eq.(\ref{poli1}), and we also
recall that ${\Delta_4}=256 D_4$. A calculation for the degenerate
cases ($I_4=0$) shows that these conditions are\beq D_4>0,\qquad
I_1^2>M,\qquad I_1I_3>M(I_1^2-M)\label{st2}\eeq\noindent yielding
the $st^2$ model and the further specialization $M=0$ results in
\beq D_4>0\qquad I_1I_3>0\label{t3}\eeq\noindent corresponding to
the $t^3$ model. Now by virtue of Eq.(\ref{explinvar}) it is easy
to see that a further specification to the case of our BPS
$D0D4-D2D6$ split renders our condition $-PvI_1<0$ equivalent to
the one $I_1I_3>0$. Hence we obtained the nice result that for
$D0D4-D2D6$ splits the conditions encapsulated in the positivity
of three-qubit
 (i.e. U-duality) invariants of  Eqs.(\ref{BPS1})-(\ref{BPS2})
 and the positivity of the four-qubit ones of
Eq.(\ref{t3}) provide a necessary condition for the consistency
condition for such two-center composites to hold.

Let us also verify explicitly that the aforementioned criteria
indeed provide real roots of our polynomial (\ref{poli1})
featuring the algebraically independent four-qubit invariants. For
the $t^3$ model one of the roots is zero due to the vanishing of
the invariant $I_4$. For the remaining three roots we have to look
at the solutions of Eq.(\ref{poly3}). After the substitution \beq
s=t-2\left(\frac{2I_1}{3}\right)\label{vissza}\eeq\noindent and
the definitions \beq {\mu }=-3\left(\frac{2I_1}{3}\right)^2,\qquad
{\nu}=2\left(\frac{2I_1}{3}\right)^3-4I_3\label{vissza2}\eeq\noindent
this cubic equation and its discriminant takes the form \beq
s^3+\mu s+\nu=0, \qquad\Delta=
\left(\frac{\mu}{2}\right)^2+\left(\frac{\nu}{3}\right)^3
\eeq\noindent where for the explicit form of $\Delta$ see
Eq.(\ref{cubicdiscr}). According to Eq.(\ref{explhyp}) $\Delta<0$,
this yields for Cardano's formula the case of "casus
irreducibilis"\cite{Bewersdorff} with explicit solutions $s_{j+1},
j=0,1,2$. Transforming back these solutions by using
Eq.(\ref{vissza}) to the variables $t_{j+1}$ we obtain the final
solutions\beq
t_{j+1}=4\left(\frac{2I_1}{3}\right)\sin^2\left(\frac{\Phi}{6}+\frac{\pi}{3}j\right),\qquad
j=0,1,2 \eeq\noindent where \beq
\sin^2\left(\frac{\Phi}{2}\right)=\frac{I_3}{\left(\frac{2I_1}{3}\right)^3}\eeq\noindent
a quantity clearly {\it positive} by virtue of
Eqs.(\ref{explinvar}) i.e. $I_1I_3>0$ for all sign combinations
compatible with Eq.(\ref{bps}). Since according to
Eqs.(\ref{cubicdiscr}) and (\ref{explhyp}) the sign of $D_4$ is
just the opposite of the sign of ${\Delta}$ we see that in the
special case of the BPS $D0D4-D2D6$ split in the $t^3$ model the
real roots we have obtained are in accord with our conditions of
Eq.(\ref{t3}) used in a more general context.

\subsection{Real roots and canonical forms}
Our explicit formulas for the real roots of the fundamental
polynomial of Eq.(\ref{poli1}) enables an explicit construction of
the canonical forms of the four qubit states associated to the
charge configurations describing two-center solutions. Let us give
just a few examples for the BPS $D0D4-D2D6$ split. First we write
the roots of the fundamental polynomial of Eq.(\ref{poli1}) for
the BPS $D0D4-D2D6$ splits of the $t^3$ model in the following
form \beq
t_{j+1}=4(e-f)\sin^2\left(\frac{\Phi}{6}+\frac{\pi}{3}j\right),\quad
j=0,1,2\qquad t_4=0\label{newsol1}\eeq\noindent \beq \cos\Phi
=\frac{e(e+3f)^2+f(f+3e)^2}{e(e+3f)^2-f(f+3e)^2},\qquad
e=\frac{Uv}{3},\qquad f=Pq.\label{newsol2}\eeq\noindent Note that
for BPS solutions we have $ef<0$. Notice also that for our special
case $2I_1=\langle {\cal Q}_1,{\cal Q}_2\rangle=3(e-f)$ cannot be
zero for BPS splits so this charge configuration is mutually
nonlocal.

As our first example let us consider the nontrivial cases $f\neq 0$, and $e\neq 0$ when $D_4=0$.
The first case is characterized by the constraint $e+3f=0$.
In this case $I_3\neq 0$ and from Eq.(\ref{newsol2}) we get $\Phi=\pi$ hence
\beq
t_1=t_3=\frac{2I_1}{3},\quad t_2=4\left(\frac{2I_1}{3}\right),\quad t_4=0,\qquad I_1=\frac{4Uv}{3}.\label{gyokok1}
\eeq
\noindent
Using now Eq.(\ref{gyokok}) up to permutations one can associate a canonical
form to this configuration as shown in Eq.(\ref{genuine}).
This situation arises for example when $U=-v=3$ and $P=q=1$.
Notice that this is a highly degenerate case since now $D_3({\cal Q})=0$,
and $W=3(1-\tau)(1+\tau^2)$, hence $Z=0$ for $\tau=\pm 1$ which is on the boundary of the lower half plane.
One can also see that for $e+3f=$ the superpotential $W$ can be written in a factorized
form.

Our second example is associated with the case when $f+3e=0$ producing the other nontrivial zero for $D_4$.
In this case $I_3=0$, and $\Phi =0$ hence we have
\beq
t_1=0,\quad t_2=t_3=3\left(\frac{2I_1}{3}\right),\quad t_4=0,\qquad
I_1=2Uv.\eeq\noindent
From Eq.(\ref{genuine}) one can see that the canonical state is proportional to the one $\vert 0101\rangle +\vert 1010\rangle$.

Our last example is  a one with $\Phi =\frac{\pi}{2}$. In this case
$e(e+3f)^2+f(f+3e)^2=0$ which is of the form
\beq
\lambda^3+5\lambda^2+\frac{5}{3}\lambda +\frac{1}{27}=0, \qquad
\lambda=\frac{Pq}{Uv}.\eeq\noindent
It is easy to check that $\lambda=-1/3$ i.e. $e+f=0$ is a solution, hence
$(\lambda +1/3)(\lambda^2+14\lambda/3 +1/9)$ is a factorized form of our polynomial, yielding the solutions
\beq
\lambda_1=-\frac{1}{3},\quad \lambda_{2,3}=\frac{1}{3}(-7\pm 4\sqrt{3}).
\eeq\noindent
Let us consider the rational solution $\lambda=-\frac{1}{3}$. In this case we get
\beq
t_1=\frac{4I_1}{3},\quad t_{2,3}=\frac{I_1}{3}(1\mp\sqrt{3})^2,\quad t_4=0,\qquad I_1=Uv.\eeq\noindent
Now the real roots are all different hence $D_4\neq 0$.
However, now $Uv=-3Pq$ hence $(Uv+Pq)(Uv+9Pq)<0$.
A consequence of this is that the consistency condition of Eq.(\ref{baromifontos}) cannot be satisfied so no charge configuration of this kind supports a two-center solution.
The canonical form can again be read off from
Eqs.(\ref{genuine})-(\ref{gyokok}).
Notice also that for all three cases $t_1+t_2+t_3+t_4=4I_1$ as it has to be, moreover the corresponding values for $\lambda$ can be written in the form $\lambda_k=-1/3^k$ for $k=0,1,2$.

It is interesting to  realize that all these special charge configurations giving rise to special four-qubit canonical forms are outside the domain of
legitimite two-center solutions. Later
when we connect the special values of $e$ and $f$ to properties of the $j$ function we have something more to say about this phenomenon.
In order to get charge configurations supporting BPS two center solutions we have to chose the asymptotic moduli from the eligible region boundend by the usual wall of marginal stability\cite{Bates}.  In our case this wall is given by the locus

\beq
4\xi x_{\infty}^2=(x_{\infty}^2+y_{\infty}^2+\xi)(x_{\infty}^2+y_{\infty}^2+\eta), \qquad \xi=\frac{U}{3P},\quad \eta =-\frac{3q}{v}.
\eeq
\noindent

\subsection{Splitting of invariants}
Notice that the quartic invariant for the $t^3$ model has the
explicit form
 \beq -D_3({\cal Q}_1+{\cal Q}_2)=
-(Uv)^2+3(Pq)^2-6UvPq+4UP^3-4vq^3. \eeq\noindent Now
$I_{+2}=4UP^3$ and $I_{-1}=-4vq^3$ hence according to
Eq.(\ref{poli2}) we can write \beq -D_3({\cal Q}_1+{\cal
Q}_2)=\Pi_4(\Lambda_{i_0i_1i_2i_3},1)=[4I_{+1}+6I_0+4I_{-1}]+I_{+2}+I_{-2}.
\eeq \noindent From this expression we see that the relationship
between the quantities $-D_3({\cal Q}_1+{\cal Q}_2)$, $-D_3({\cal
Q}_{1,2})$  is governed by the combination $4I_{+1}+6I_0+4I_{-1}$
of  $SL(2)_0$ {\it covariants}. Such relationships are needed for
studying situations when the two-center solution is BPS, but the
corresponding single center one is not.

For $D0D4-D2D6$
splits the quantity above can also be written in two equivalent forms featuring
quantities related to factors of $SL(2)_0$ {\it invariants}
\beq
-D_3({\cal Q}_1+{\cal Q}_2)=-8UvPq-2I_1(Uv+Pq)-D_3({\cal
Q}_1)-D_3({\cal Q}_2),\eeq\noindent or\beq -D_3({\cal Q}_1+{\cal
Q}_2)= -24(Pq)^2-2I_1(Uv+9Pq)-D_3({\cal Q}_1)-D_3({\cal Q}_2)
 .\eeq\noindent
Hence if both of the constituents are BPS i.e. $-D_3({\cal
Q}_1)>0$ and $-D_3({\cal Q}_2)>0$, the conditions for the single
centered system to be BPS too i.e. $-D_3({\cal Q}_1+{\cal Q}_2)>0$
are again governed by a four qubit invariant $I_1$ or by the factors
of another four qubit one namely $D_4$. It is amusing to realize
that all of the factors of $D_4$ are appearing in these two
equations.

As an example one can see that if $I_1<0$ and $D>0$ and moreover
$Uv+Pq>0$ (which is a factor of $D_4$), then two BPS
configurations again yield a BPS one. In the example of Bates and
Denef\cite{Bates} $U=-v=4$ and $P=q=1$, $-D_3({\cal Q}_1)>0$ and
$-D_3({\cal Q}_2)>0$, hence $D_4>0$ and $I_1<0$ but $Uv+Pq<0$
hence $-D_3({\cal Q}_1+{\cal Q}_2)=-125<0$. In this case a single
center BPS solution does not exist, although the corresponding
two-center one does.

\subsection{The $j$ invariant}

It should be ovious by now that the basic mathematical object
giving rise to our observations is the fundamental polynomial of
Eq.(\ref{poli1}) featuring all of our algebraically independent
four-qubit invariants. For no matter what kind of two center
charge configuration we have we can construct this quartic
polynomial. As a next step we can calculate its resolvent qubic of
Eq.(\ref{rcubic}). After the substitution $u=2x$ this polynomial
takes the form $2(4x^3-Sx-T)$. As a next step to this resolvent cubic we
associate an
elliptic curve of Weierstrass canonical form as \beq
y^2=4x^3-Sx-T.\label{elliptic}\eeq\noindent The $j$ invariant of
this curve is defined as
 \beq
j=1728\frac{S^3}{S^3-27T^2}.\label{jinvariant}\eeq\noindent In
particular for the special case of the $t^3$ model we get \beq
j=\frac{(2I_1)^3\left((2I_1)^3-24I_3\right)^3}
{I_3^3\left((2I_1)^3-27I_3\right)} .\eeq\noindent After recalling
the definitions of Eq.(\ref{newsol2}) a further specification to
our case of the BPS $D0D4-D2D6$ split one obtains \beq
1728S^3=(2I_1)^3((2I_1)^3-24I_3)^3=27(e-f)^3[27e(3f+e)^2-3f(3e+f)^2]^3,\eeq\noindent
\beq 256D_4=S^3-27T^2=-27[e(3f+e)^2][f(3e+f)^2]^3.\eeq\noindent
Now  \beq
27e(3f+e)^2-3f(3e+f)^2=(3e-\xi_1f)(3e-\xi_2f)(3e-\xi_3f)\eeq\noindent
where \beq \xi_{j+1}=-5+4\sqrt[3]{2}\omega^j,\qquad
\omega_j=e^{2\pi j/3},\qquad j=0,1,2.\eeq\noindent Interestingly
the number $\xi_1$ can also be written in the form \beq
\xi_1=-5+4\sqrt[3]{2}=3\left(\frac{1-\sqrt[3]{2}}{1+\sqrt[3]{2}}\right)^2
\label{valoslesz}\eeq\noindent moreover, one can check that
$\xi_1+\xi_2+\xi_3=-15$, $\xi_1\xi_2+\xi_2\xi_3+\xi_1\xi_3=75$ and
$\xi_1\xi_2\xi_3=3$.

Finally one obtains for the $j$ invariant the expression \beq
j=\frac{(f-e)^3}{[e(3f+e)^2][f(3e+f)^2]^3}[(3e-\xi_1f)(3e-\xi_2f)(3e-\xi_3f)]^3.
\eeq\noindent
 Clearly the denominator of $j$ is zero precisely when $D_4$ is
vanishing. According to the results of the previous subsections
for BPS ($ef<0$) composites when this happens the consistency
condition cannot be satisfied. The numerator on the other hand can
vanish either when $2I_1=\langle{\cal Q}_1,{\cal Q}_2\rangle=0$
i.e. the charge configuration is local, or when the remaining real
factor vanishes i.e. when $Uv=\xi_1Pq$. Since according to
Eq.(\ref{valoslesz}) $\xi_1$ is positive, for BPS composites the
latter two conditions cannot be satisfied. Hence the $j$ function
is not having any pathological behavior for the physically
legitimate cases of BPS $D0D4-D2D6$ composites.

Notice however, that the $j$ function can be made to vanish for
non-BPS $D0D4-D2D6$ composites. In this case we have $ef>0$, and
one can try to combine the elementary non-BPS solutions obtained
for the $D0D4$ and $D2D6$ systems\cite{Soroush}. For such non-BPS
composites with {\it local} charge  configurations a
generalization of the consistency condition has recently been
given\cite{Bena}.

\section{Elliptic curves}

\subsection{Elliptic curves and four-qubit states}
Let us try to justify that our association of an elliptic curve
to the resolvent cubic Eq.(\ref{rcubic}) of our fundamental
polynomial of Eq.(\ref{poli1}) is a natural one. For this it is
useful to regard our three-qubit charge states $\vert{\cal
Q}_1\rangle$ and $\vert{\cal Q}_2\rangle$ with amplitudes given by
Eqs.(\ref{charge}),(\ref{atteres3})-(\ref{atteres4}) as the ones
embedded into more general unnormalized four-qubit ones with {\it
complex} amplitudes. (We will have something more to say about
this embedding later.) Hence we can use the algebraically closed
field of complex numbers to put our considerations in a more
general setting.

Let us now notice that according to Luque and Thibon\cite{Luque}
we can write our key quantity $256D_4=S^3-27T^2$ in yet another
form by rewriting the polynomial invariants $S$ and $T$ as
\beq 12S=U^2-2V,\qquad
216T=U^3-3UV+216D^2\label{Lu1}\eeq\noindent where \beq
U=H^2+4(M-L),\qquad V=12(HD-2LM).\label{Lu2}\eeq\noindent The
algebraically independent invariants\cite{Luque} $H,L,M,D$ showing
up in these expressions are related to our set $I_1,I_2,I_3,I_4$
as\cite{Levay4} \beq I_1=\frac{1}{2}H,\qquad
I_2=\frac{1}{6}(H^2+2L+4M),\qquad I_3=D+\frac{1}{2}HL,\qquad
I_4=L.\label{relatedinv}\eeq\noindent

It is known that a nonsingular projective curve of genus $1$ is
isomorphic to a plane cubic curve of the Tate form\cite{Ueno} \beq
x_0x_2^2+a_1x_0x_1x_2+a_3x_0^2x_2-x_1^3-a_2x_0x_1^2-a_4x_0^2x_1-a_6x_0^3=0.
\label{elli}\eeq\noindent Here the coefficients
$a_1,a_2,a_3,a_4,a_6$ can be taken from an algebraically closed
field ${\mathbb F}$ of arbitrary characteristic and
$(x_0,x_1,x_2)$ are homogeneous coordinates of the projective
plane ${\mathbb F}P^2$. (Of course for qubits we are merely
interested in the complex case i.e. our concern will be curves in
${\mathbb C}P^2$.) Using inhomogeneous coordinates $(x,y)\equiv
(x_1/x_0,x_2/x_0)$ this can be rewritten as \beq
y^2+a_1xy+a_3y=x^3+a_2x^2+a_4x+a_6.\label{elli2}\eeq\noindent Now
the $j$ invariant and discriminant $\Delta$ of a plane cubic curve
of this form is defined\cite{Ueno} as \beq
j=\frac{c_4^3}{\Delta}=1728\frac{c_4^3}{c_4^3-c_6^2}\eeq\noindent
where \beq c_4=b_2^2-24b_4,\qquad
c_6=-b_2^3+36b_2b_4-216b_6\label{eh1}\eeq\noindent \beq b_2=a_1^2+4a_2,
\qquad b_4=a_1a_3+2a_4,\qquad b_6=a_3^2+4a_6.\label{eh2}\eeq\noindent A
necessary and sufficient condition for our curve to be nonsingular
is $\Delta\neq 0$.

Now looking at Eqs.(\ref{Lu1})-(\ref{Lu2}) one can easily check
that four-qubit states provide a particularly nice parametrization
for a special class of plane cubic curves with $a_6\equiv 0$.
Indeed setting \beq b_2=U=H^2+4(M-L),\qquad
b_4=\frac{1}{12}V=HD-2LM,\qquad b_6=D^2\eeq\noindent yields \beq
c_4=12S,\qquad c_6=-216T.\eeq\noindent Moreover, one can also see
that \beq a_1=H=2I_1,\qquad a_2=M-L,\qquad a_3=D=I_3-I_1I_4,\qquad
a_4=-LM,\qquad a_6=0\eeq\noindent hence the family of such cubic
curves for the stu model $(H\neq 0,D\neq 0,M\neq 0,L\neq 0 )$ is \beq
y^2+Hxy+Dy=x^3+(M-L)x^2-LMx\label{stucurve}\eeq\noindent for the
$st^2$ model ($H\neq 0,D\neq 0,M\neq 0,L=0$)  \beq
y^2+Hxy+Dy=x^3+Mx^2\label{st2curve}\eeq\noindent and finally for
the $t^ 3$ model ($H\neq 0,D\neq 0,L=0,M=0$) it is\beq
y^2+Hxy+Dy=x^3.\label{t3curve}\eeq\noindent

For all of these curves the hyperdeterminant is given by the usual
formula
 $256D_4=S^3-27T^2$ with the $j$ invariant having the form of Eq.(\ref{jinvariant}).
As an example let us take a look at the hyperdeterminant $D_4$ for
the $st^2$ model which is given by the expression \beq
256D_4=I_3^2\left(I_3[(2I_1)^2-27I_3]+36M[I_2^2-2I_1I_3]\right),\qquad
2M=3I_2-2I_1^2.\label{hyphyp}\eeq\noindent One can see that in accordance with
our association of cubic curves to the stu model and its
truncations for $M=0$ we get back to Eq.(\ref{explhyp}) i.e. the
expression for the $t^3$ model.

Recall also that if the characteristic of the field is not $2$ or
$3$ via further projective transformations (i.e. completing the
square on the left hand side  and then the cube on the right hand
side of Eq.(\ref{stucurve})) one obtains the Weierstrass canonical
form of Eq.(\ref{elliptic}) which is simply related to the
resolvent cubic of our polynomial $\Sigma_4$ of Eq.(\ref{poli1})
we have started our considerations with. Notice however, that
although due to the fact that now ${\mathbb F}={\mathbb C}$ the
Weierstrass form in our case can always be reached, this form is
not showing the basic differences between the structures of the
cubic curves associated to the $stu$, $st^2$ and $t^3$ models,
hence we prefer the Tate form.

In the following as usual by an elliptic curve we will mean a
nonsingular plane cubic curve as given by Eq.(\ref{elli}) and  the
"point at infinity" $(x_0,x_1,x_2)=(0,0,1)$. Since for these
curves we have only $(0,0,1)$ as a point at infinity by an abuse
of notation one can call Eq.(\ref{elli2}) as the defining equation
for the corresponding elliptic curve. An elliptic curve is defined
over a subfield ${\mathbb F}^{\prime}$ of  ${\mathbb F}$ if all
the coefficients $a_1,a_2,a_3,a_4,a_6$ are taken from ${\mathbb
F}^{\prime}$ .
In the case of the discrete version of the continuous $U$-duality group,
as nonperturbative string symmetries, the coefficients of this curve
are the algebraically independent four-qubit invariants $H,L,M,D$
(related to the ones $I_1,I_2,I_3,I_4$ via Eq.(\ref{relatedinv}))
taken from the ring of integers of the subfield ${\mathbb Q}$.
In this case the arithmetic aspects of the theory of elliptic curves
within the context of two-center black hole solutions becomes important\cite{Moore}.

The upshot of these considerations is that to a
particular charge configuration of two-center black holes of the
$stu$ model we can indeed associate quite naturally an elliptic
curve of {\it special kind} (i.e. $a_6=0$). This elliptic curve is
of the form of Eq.(\ref{stucurve}).

Notice that the invariants $H,L,M$ play an important physical
role. Since $H=\langle{\cal Q}_1,{\cal Q}_2\rangle$ the vanishing
of this invariant gives rise to local charge configurations, in
this case the $xy$ term of Eq.(\ref{stucurve}) is missing. The
vanishing conditions on $L$ and $M$ give rise to truncations to
the $st^2$ and $t^3$ models.
Notice also that since for $L=0$ we have $D=I_3$ hence in the $st^2$ and $t^3$
models sending $D$ to zero corresponds to taking the
 limit when our elliptic curve degenerates (see Eqs.(\ref{explhyp}), (\ref{relatedinv}) and (\ref{hyphyp})).

In establishing the correspondence between two center charge
configurations and elliptic curves we tacitly assumed that our
unnormalized charge states with real amplitudes (transforming
according to the corresponding representation of the {\it
continuous} $U$-duality group valid in the supergravity
approximation) are regarded as ones embedded in unnormalized
four-qubit states with complex amplitudes. However, until this
point we did not specify what kind of complex states we are having
in mind.

 In a recent series of papers it was shown\cite{Levay3,Levszal}
that for single center extremal stationary spherically symmetric
BPS and non-BPS black hole solutions apart from {\it charge
states} i.e. unnormalized three-qubit states with {\it real}
amplitudes, we can also introduce unnormalized {\it complex}
three-qubit states which are depending on the charges and also on
the moduli fields. (For a more general class of such states even the warp factor can be included\cite{Levszal}.) It has been shown that the
complex amplitudes $\psi_{000},\psi_{001},\dots,\psi_{111}$ of
such states $\vert\psi\rangle$ are related to the central charge
and its covariant derivatives with respect to the moduli. For
example the amplitudes featuring the central charge have the
form\cite{Levay3} \beq {\psi}_{000}=\sqrt{2}\overline{Z},\qquad
\psi_{111}=-\sqrt{2}Z.\eeq\noindent Moreover, recently
it has also been demonstrated that such complex three-qubit states
are just special types of four-qubit 
ones\cite{Levayfour}. It is natural to expect that for two center
solutions complex four-qubit states of similar type should show
up. For example in the consistency condition of
Eq.(\ref{necessary}) the term ${\rm Im}(\overline{\cal Z}_2{\cal
Z}_1)$ can be reinterpreted as a truncation of a pairing between
the two complex amplitudes $\Psi_{0000}\equiv \sqrt{2}{\cal Z}_1$
and $\Psi_{1111}=-\sqrt{2}\overline{\cal Z}_2$ of such four-qubit
states. In this respect recall the formalism developed by
Denef\cite{Denef2} in the type IIB duality frame where apart from
the usual antisymmetric, topological moduli independent
intersection product (our four-qubit invariant $I_1$ in the $stu$
model) the importance of the symmetric, positive definite, moduli
dependent Hodge product is emphasized (in the $stu$ model giving
rise to the interpretation of the black hole potential as the norm
of a three-qubit state\cite{Levay3}) . Such structures are the
natural ones appearing in the multicenter black hole context. In
the $stu$ case it is easy to establish  a correspondence between
this formalism and the one where {\it complex} three and
four-qubit states appear\cite{levnext}.
We conjecture that such
moduli dependent complex states and their associated invariants
featuring elliptic curves could be the relevant mathematical
objects hiding behind the considerations of this paper based merely on charge states.
Note that moduli dependent curves of that kind would also display
explicit dependence on extra complex parameters giving rise to elliptic fibrations. We will
have something more to say about this possibility in Section V.

\subsection{Degeneracies}
The discriminant ${\Delta}$ of our elliptic curve is related to
the hyperdeterminant $D_4$ as $256D_4=\Delta$. For nonvanishing
$D_4$ the curves are nonsingular. Since our curves are of genus
$1$ they are topologically tori. Vanishing of $D_4$ results in
different degeneracies of these tori (for example one of their
homologically nontrivial cycles can contract to a point). One can
illustrate this in the $stu$ and
$st^2$ models when both of the two BPS constituents are having
vanishing Bekenstein-Hawking entropy i.e $I_{\pm 2}=0$ for both
centers.

Indeed, let us use the parametrization as introduced by
Sen\cite{Sen2}
for such two-center composites ${\cal Q}={\cal Q}_1+{\cal Q}_2$

\beq
\begin{pmatrix}Q\\P\end{pmatrix}=
\begin{pmatrix}adQ-abP\\cdQ-bcP\end{pmatrix}+
\begin{pmatrix}abP-bcQ\\adP-cdQ\end{pmatrix},\qquad
\begin{pmatrix}a&b\\c&d\end{pmatrix}\in SL(2,{\mathbb Z}).
\label{sendecomp}
\eeq
\noindent
Here $P$ and $Q$ are four-vectors related to the ones of Eq.(\ref{atteres1})-(\ref{atteres4})
as follows
\beq
A=a(dQ-bP),\quad B=c(dQ-bP),\quad C=b(aP-cQ),\quad D=d(aP-cQ).
\eeq
\noindent
Since the vectors $A$ and $B$ and $C$ and $D$
are now proportional the bivectors of $X$ and $Y$ of
Eq.(\ref{bivectors}) are zero hence $I_{\pm 2}=I_{\pm 1}=0$ though $I_0\neq
0$. Now Eq.(\ref{Suj}) shows that
\beq S=3I_0^2,\qquad
T=I_0^3,\label{senst}\eeq\noindent
hence  $S^3-27T^2=0$ i.e. $D_4=0$.
The nonseparable bivector is $Z=\frac{1}{2}Q\wedge P$ hence
\beq
I_0=\frac{1}{3}Z^2=\frac{1}{12}(Q\wedge P)\cdot(Q\wedge P)=-\frac{1}{6}D_3({\cal Q}).
\label{i0}
\eeq
\noindent
Alternatively one can directly check that $I_3=I_4=0$ and
$6I_2=-6I_0=D_3({\cal Q})$ in accordance with Eq.(\ref{kellenifog}), hence $S=3I_2^2$
and $T=I_2^3$ yielding again $D_4=0$.

In order to obtain the Tate form of our elliptic curve we have to
calculate the quantities $H,L,M,D$. One can show that \beq
H=\hat{P}\cdot\hat{Q},\qquad L=0,\qquad
M=-\frac{1}{4}{\hat{P}}^2{\hat{Q}}^2,\qquad
D=0\label{seninv}\eeq\noindent where we introduced the notation
\beq
\begin{pmatrix}\hat{Q}\\\hat{P}\end{pmatrix}=\begin{pmatrix}d&-b\\-c&a\end{pmatrix}
\begin{pmatrix}Q\\P\end{pmatrix}.\label{senm}\eeq\noindent
Now the elliptic curve of Eq.(\ref{stucurve}) is of the form \beq
y^2+
\hat{P}\cdot\hat{Q}xy=x^3-\frac{1}{4}{\hat{P}}^2{\hat{Q}}^2x^2.\label{Tatesen}\eeq\noindent
Completing the square on the left hand side (and by an abuse of
notation using $y$ for the new variable again) yields \beq
y^2=x^2\left(x+\frac{1}{4}D_3({\cal Q})\right).
\label{nbpscurve}\eeq\noindent Notice that the quantities
$\hat{P}$ and $\hat{Q}$ hence the Tate form of the cubic curve is
featuring the parameters $a,b,c,d$ characterizing the two-center
split, however the latter form is not. This is due to the fact
that $D_3({\cal Q})=(P\cdot Q)^2-P^2Q^2 = (\hat{P}\cdot
\hat{Q})^2-{\hat{P}^2}{\hat{Q}^2}$. This leads us to the important
observation that at least for this singular example two-center
charge configurations labelled by elements of $SL(2,{\mathbb Z})$
correspond to the same singular curve with the (\ref{nbpscurve})
canonical form . Hence apart from the four copies of $SL(2)$s
related to four-qubit systems there seems to be a {\it fifth}
$SL(2)$ parametrizing a hidden torus. We will return to this
important point later.

Eq.(\ref{nbpscurve}) represents a cubic curve with a {\it node} at
the point $(0,0)$. The tangent lines at $(0,0)$ are obtained from
the equation $y^2-(D_3/4)x^2=0$. Such lines are of the form \beq
\left(y-\frac{1}{2}\sqrt{D_3}\right)
\left(y+\frac{1}{2}\sqrt{D_3}\right)=0.\label{lines}\eeq\noindent

It is important to recall that we assumed that all of our curves
are over the complex numbers, i.e. ${\mathbb F}={\mathbb C}$.
Hence for a cubic in the canonical form $y^2=x^3+\alpha x+\beta$
we are still free to use transformations of the form $(x,y)\mapsto
(\mu^2 x, \mu^3 y)$ and
$(\alpha,\beta)\mapsto(\mu^4\alpha,\mu^6\beta)$, where $\mu\in
{\mathbb C}^{\times}$. Using the transformation $(x,y)\mapsto
(-x,-iy)$, i.e. $\mu=i$ transforms Eq.(\ref{nbpscurve}) to the
form \beq y^2=x^2\left(x-\frac{1}{4}D_3({\cal Q})\right).
\label{bpscurve}\eeq\noindent

As we see such transformations are similar to the ones changing a BPS charge configuration ($D_3<0$)
to a non-BPS one
($D_3>0$). Interestingly as shown by Eq.(\ref{lines})
 (and a similar one obtained from Eq.(\ref{bpscurve}))
 the pair of lines at the node are parametrized by $\sqrt{-D_3}$
 for the BPS-case or $\sqrt{D_3}$ for the non-BPS one,
 i.e. quantities proportional to the Bekenstein-Hawking entropy.

Notice also that Eq.(\ref{bpscurve}) can also be parametrized as
\beq
y^2=3t(x^{\prime}-t)^2+(x^{\prime}-t)^3=(x^{\prime})^3-3t^2(x^{\prime})-2t^3,\qquad t=-\frac{D_3}{12}
\eeq
\noindent
Using Eqs.(\ref{senst})-(\ref{i0}) we can write this as
\beq
y^2={x^{\prime}}^3-\frac{1}{4}S{x^{\prime}}-\frac{1}{4}T\label{redukalni}.\eeq\noindent
Now a further transformation $(u,v)=(x/2,x/8)$ i.e. $\mu^2=1/2$ transforms this
to the form $v^2=u^3-Su-2T$ of Eq.(\ref{rcubic}), with the left hand side being just the resolvent cubic of $\Sigma_4=0$.
The procedure discussed here illustrates how the general Tate form of our cubic curve of Eq.(\ref{Tatesen})
can be transformed to a resolvent cubic form.

It is important to realize that if the domain of definition of our
curve is the subfield ${\mathbb Q}$ from the general Tate form of
Eq.(\ref{stucurve}) merely the form \beq
y^2=x^3-\frac{1}{4}Sx+\frac{1}{4}T \label{kifuto} \eeq\noindent
can be reached. Further reduction is possible only if $\mu$ is a
square in ${\mathbb Q}$. In the case of our singular curve of
Eq.(\ref{nbpscurve}) if $D_3$ is not a square we cannot represent
the tangent lines in the form of Eq.(\ref{lines}) hence the
tangent lines in this case are not rational.

Due to permutation symmetry we have no parametrization of
(\ref{sendecomp}) type for the $t^3$ model available. However, as
a degenerate example in this case we can consider the example of
Bates and Denef\cite{Bates} instead by sending one of the charges
e.g. $P$ to zero. In this case one of the black holes is a small
one, and the two center composite is characterized by the charges
$U,v,q$. Now one checks that $I_3=0$, hence according to
Eq.(\ref{explhyp}) we get again $D_4=0$. The explicit form of the
corresponding degenerate elliptic curve is \beq y^2+Hxy=x^3,
\qquad H=\langle{\cal Q}_1,{\cal
Q}_2\rangle=Uv,\label{j0}\eeq\noindent with $j=\infty$ just like
in the previous example.

\subsection{Invariance properties}

Since the coefficients $a_j$ of the elliptic curve of Eq.(\ref{stucurve}) are invariants the same curve is associated to equivalent two-center charge configurations in the $stu$ model.
In other words our curve is clearly invariant under the action of the group $SL(2,{\mathbb R})_0\times G_4$, where $SL(2,{\mathbb R})$ is the horizontal symmetry group of generalized exchange transformations
of the centers and $G_4$ is the $d=4$ continuous $U$-duality group of the $stu$ model which is $SL(2,{\mathbb R})^{\times 3}$.

However, as far as physics is concerned in a special case of the
$t^3$ model we have also found some connection between the
structure of $SL(2,{\mathbb R})_0\times G_4$ (four-qubit)
invariants and the structure of the consistency condition of
Eq.(\ref{necessary}). This connection at first sight  is not
surprising since the consistency condition contains a four-qubit
invariant $2I_1=\langle{\cal Q}_1,{\cal Q}_2\rangle$, however more
importantly it is also featuring the quantity ${\rm
Im}(\overline{\cal Z}_2{\cal Z}_1)$ which is not invariant with
respect to the full group of $SL(2,{\mathbb R})_0\times  G_4$
transformations.

In order to show this let us observe however, that this quantity
is invariant under transformations belonging to the horizontal
subgroup $SL(2,{\mathbb R})_0$. For let us recall our definition
of Eq.(\ref{charge}) and then write the central charges as \beq
{\cal
Z}_1=e^{K/2}(Q_0+Q_1\tau_1+Q_2\tau_2+Q_3\tau_3-P^1\tau_2\tau_3
-P^2\tau_1\tau_3-P^3\tau_1\tau_2+P^0\tau_1\tau_2\tau_3) \eeq
\noindent \beq {\cal
Z}_2=e^{K/2}(q_0+q_1\tau_1+q_2\tau_2+q_3\tau_3-p^1\tau_2\tau_3
-p^2\tau_1\tau_3-p^3\tau_1\tau_2+p^0\tau_1\tau_2\tau_3), \eeq then
for $a,b,c,d\in{\mathbb R}$ we have ${\cal Z}^{\prime}_1=a{\cal
Z}_1+b{\cal Z}_2$, and $\overline{\cal
Z}^{\prime}_2=c\overline{\cal Z}_1+d\overline{\cal Z}_2$, hence
${\rm Im}(\overline{\cal Z}^{\prime}_2{\cal Z}^{\prime}_1)= {\rm
Im}(\overline{\cal Z}_2{\cal Z}_1)$ due to $ad-bc=1$. However,
this quantity is invariant merely under a special subgroup\cite{Manschot} of
$G_4$. Indeed, we have \beq {\cal Z
}(K\tau_1,K\tau_2,K\tau_3;K{\cal Q}_i)= {\cal Z
}(\tau_1,\tau_2,\tau_3;{\cal Q}_i) \eeq \noindent where $K$ is the
subgroup of $G_4$ transformations of the form \beq
\begin{pmatrix}1&0\\k_1&1\end{pmatrix}\otimes
\begin{pmatrix}1&0\\k_2&1\end{pmatrix}\otimes\begin{pmatrix}1&0\\k_3&1\end{pmatrix}\vert{\cal Q}_i\rangle,\qquad \tau_a\mapsto \tau_a+k_a\qquad k_a\in{\mathbb R},\quad a=1,2,3.
\eeq \noindent Notice that the group $K$ with $k_a\in{\mathbb Z}$
is precisely the stabilizer of the cusps in the three copies of
the fundamental domain of the modular group, i.e. the stabilizer
of $\tau_a=-i\infty$ for $a=1,2,3$.

Since the walls of marginal stability are determined by the
central charges and for the four qubit invariant $2I_1$ we have
$\langle K{\cal Q}_1,K{\cal Q}_2\rangle=\langle{\cal Q}_1,{\cal
Q}_2\rangle$ clearly wall-crossing does not obstruct the $K$
subgroup. Therefore we see that unlike our elliptic curves the
consistency condition is left invariant (in the above sense)
merely with respect to the subgroup $SL(2,{\mathbb R})_0\times K$.
Hence the correspondence between the physics of two center
solutions and our special class of elliptic curves should be
refined (see in this respect the comments at the end of Section
V.A).

Now we turn to another important issue we have not discussed yet.
Based on our experience with the degenerate case studied in the
previous subsection we expect that for the nondegenerate  case our
mapping of two center charge configurations characterized by
four-qubit invariants to elliptic curves should be many to one.
Moreover, since in the Tate form we should have $a_6=0$ our
mapping in the $stu$ case cannot be onto either. We also know from
the theory of elliptic functions that if we work in the algebraic
closure of our field ${\mathbb F}$  i.e. ${\mathbb F}^{\rm alg}$
two elliptic curves are isomorphic if and only if their $j$
invariants are the same. Hence in the case of ${\mathbb R}$
working with the complex numbers charge configurations with {\it
different} four-qubit invariants could be mapped to isomorphic
elliptic curves with the same $j$ invariant. In the singular case
according to Eqs.(\ref{seninv})-(\ref{senm}) the values of $H$ and
$M$ are clearly different for different splits, however $j=\infty$
in all cases.

Now over ${\mathbb C}$ every elliptic curve $E$ is isomorphic to
an elliptic curve $E(\Lambda)$ where $\Lambda$ is a lattice in the
complex plane and the mapping between ${\mathbb C}/\Lambda$ (a
torus) and $E(\Lambda)$ is an isomorphism provided by the usual
map defined by the Weierstrass ${\cal P}$ function. Explicitly
$E(\Lambda)$ is of the form \beq
x_0x_2^2=4x_1^3-g_2(\hat{\tau})x_1x_0^2-g_3(\hat{\tau})x_0^3
\label{eisenstein}\eeq\noindent where
$g_2(\hat{\tau})=60G_4(\hat{\tau})$ and
$g_3(\hat{\tau})=140G_6(\hat{\tau})$ with $G_{2k}(\hat{\tau})$ are
the Eisenstein series \beq G_{2k}(\hat{\tau})=
\sum_{(m,n)\in{\mathbb Z}^2,(m,n)\neq
(0,0)}\frac{1}{(m+n{\hat{\tau}})^{2k}} \label{eisenstein2}\eeq
\noindent and if $x\equiv x_1/x_0$ and $y=x_2/x_0$ then $x={\cal P}(z)$ and $y={\cal P}^{\prime}(z)$. Here the
lattice vectors of $\Lambda$ in ${\mathbb C}$ are $\omega_1=1$ and
$\omega_2=\hat{\tau}$ with $\hat{\tau}\in{\mathbb H}$ being the
modular parameter of the torus and the Weierstrass ${\cal P}$
function is defined as \beq {\cal
P}(z)=\frac{1}{z^2}+\sum_{(m,n)\in{\mathbb Z}^2,(m,n)\neq
(0,0)}\left[\frac{1}{(z+m+n\hat{\tau})^2}-\frac{1}{(m+n\hat{\tau})^2}\right].\eeq
\noindent Here we have used the notation $\hat{\tau}$ for the
modular parameter in order not to confuse it with the moduli
$\tau_1,\tau_2,\tau_3$ of the $stu$ and with the moduli $\tau$ of
the $t^3$ model. Now the $j$ invariant gives rise to the $j$
function \beq
j(\hat{\tau})=1728\frac{g_2(\hat{\tau})^3}{g_2(\hat{\tau})^3-27g_3(\hat{\tau})^2}.
\eeq\noindent Comparing this with the expression of the $j$
invariant as given in terms of the four-qubit invariants $S$ and
$T$ of Eq.(\ref{jinvariant}) to a two center charge configuration
we can associate a torus with modular parameter $\hat{\tau}$. Now
$j:{\mathbb H}\to {\mathbb C}$ is an automorphic function i.e.
\beq
j\left(\frac{a\hat{\tau}+b}{c\hat{\tau}+d}\right)=j(\hat{\tau}),\qquad
\begin{pmatrix}a&b\\c&d\end{pmatrix}\in SL(2,{\mathbb Z}).\label{modprop}\eeq\noindent
It can be regarded\cite{Kra} as a map providing a holomorphic
universal covering of the orbifold ${\cal H}/PSL(2,{\mathbb Z})$
(i.e. the fundamental domain for $PSL(2,{\mathbb Z})$). Its
Fourier series at $\hat{\tau}=i\infty$ is provided by the variable $u\equiv
\exp{2\pi i\hat{\tau}}$ as \beq
j(\hat{\tau})-744=\frac{1}{u}+\sum_{n=1}^{\infty}c_nu^n,\qquad
c_n\in {\mathbb Z}^{+}\cup\{0\}.\eeq\noindent From this we see
that the two-center charge configurations with discriminant
$D_4=0$ of the previous subsection with $j=\infty$ should
correspond to the modular parameter $\hat{\tau}=i\infty$ i.e. the
cusp of the Riemann surface ${\cal H}/PSL(2,{\mathbb Z})$. The
different possible two-center splits labelled by elements of
$SL(2,{\mathbb Z})$ not changing the value of the $j$ invariant
seem to be related to the modular property of the $j$-function as
shown by Eq.(\ref{modprop}). It is also intriguing to recall that
the tangent lines at the node of the degenerate elliptic curve are
parametrized by the Bekenstein-Hawking entropy (see
Eq.(\ref{lines})).

From these investigations it is natural to conjecture that similar
invariance properties of other two-center splits with $D_4\neq 0$
should hold. It would be nice to uncover the physical role of this
hidden torus and the extra $SL(2)$ symmetry associated with it.
Some speculations on how the extra modular parameter $\hat{\tau}$
should be implemented into a four-qubit picture will be given
in Section VI.

\subsection{A conjecture}

Let us consider the elliptic curve of Eq.(\ref{stucurve}) associated to two-center charge configurations of the $stu$ model.
As we have noticed this curve is of the Tate form satisfying the special constraint $a_6=0$.
The coefficients $a_j$ are four-qubit polynomials with definite degree of homogeneity $2j$. Indeed, $a_1=H$ is a quadratic, $a_3=D$ is a sextic,
$a_2=M-L$ is a quartic, and $a_4=-LM$ is an octic polynomial in the amplitudes
of the four-qubit charge state. (The variables $x$ and $y$ should be assigned the degrees four and six respectively.) Now according to this observation the constraint $a_6=0$ should be arising from the vanishing condition of a polynomial of degree $12$.
It is easy to identify this polynomial constraint. It is just the constraint\cite{Ferretal}  ${\cal P}_{12}=0$  of Eq.(\ref{p12}), or alternatively of Eq.(\ref{jolvan1}), hence in a redundant notation  for the $stu$ model we can write
\beq
y^2+Hxy+Dy=x^3+(M-L)x^2-LMx+{\cal P}_{12}.
\label{stucurve2}
\eeq
\noindent

Since the $stu$ model can be regarded as a consistent truncation of $N=8$, $d=4$ maximal supergravity with continuous\cite{Ferretal2} $U$-duality group $G_4\equiv E_{7(7)}$ one might conjecture that it should be possible to substantially generalize our considerations concerning two-center charge configurations  by studying elliptic curves of the form
\beq
y^2+P_2yx+P_6y=x^3+P_4 x^2+P_8x+P_{12}
\label{magiccurve}\eeq\noindent
where $P_{2j}$ are polynomial invariants of the group
$SL(2)_0\times G_4$ where $SL(2)_0$ refers to the horizontal symmetry group\cite{Ferretal} acting as a generalized exchange symmetry group on the two centers.
Now we have two sets of $56$ component charge vectors where each of these vectors is transforming according to the symplectic irreducible representation of $G_4$ which is now the fundamental of $E_{7(7)}$.
According to an analysis of two centered magical charge orbits\cite{Ferretal2}
working in the complexification of $G_4$ one discovers that the two centered charge orbits correspond to different real forms of the quotient of the complex groups $E_7/SO(8)$. In particular we have two $\frac{1}{8}$-BPS two-centered charge orbits with one of them being $E_{7(7)}/SO(4,4)$.
Recall now Eq.(\ref{isom}) displaying in its structure the group theoretical
reason for our occurrence of four-qubit states.
According to this equation one should be able to obtain our four-qubit invariants as special cases of $SL(2)_0\times G_4$ ones.
Hence we conjecture that an elliptic curve of the (\ref{magiccurve}) form
should display this reduction procedure via  implementing the constraint
${\cal P}_{12}=0$ via a suitable reduction of some unknown $SL(2)_0\times G_4$ polynomial invariant
$P_{12}$.

In fact many pieces of such invariants are already at our disposal.
For example the $N=8$ analogues $J_{0},J_{\pm 1},J_{\pm 2}$ of the four-qubit covariants $I_0,I_{\pm 1},I_{\pm 2}$ of Eq.(\ref{inv1}) arising from the polarization of Cayley's hyperdeterminant can now be obtained from the polarization of Cartan's quartic invariant\cite{Ferretal2}.
Using these explicit expressions one can construct the quantities
answering the similar ones of Eq.(\ref{Suj})
\beq
{\cal S}=3J_0^2-4J_{+1}J_{-1}+J_{+2}J_{-2},\qquad
{\cal T}=J_0^3+J_{+1}^2J_{-2}+J_{-1}^2J_{+2}-J_{+2}J_0J_{-2}-2J_{+1}J_{0}J_{-1}.
\label{Suj2}\eeq\noindent
Indeed precisely these polynomials of order $8$ and $12$ has been suggested as
obvious candidates for members of a  complete basis for $SL(2)_0\times G_4$
invariants.
Now one can define an elliptic curve
\beq
y^2=x^3-\frac{1}{4}{\cal S}x+\frac{1}{4}{\cal T}
\label{befuto}
\eeq
\noindent
as the one corresponding to Eq.(\ref{kifuto}).
As was commented there this form can always be obtained from the Tate form
if the characteristic of the field is neither $2$ nor $3$.
Hence we conclude that our Tate form Eq.(\ref{magiccurve}) featuring the unknown invariants $P_{2j}$  should reduce to Eq.(\ref{befuto}) after completing the square on the left and the cube on the right hand side.
Explicitly this process amounts to the transition from using the coefficients $a_j=P_{2j}$
to using the ones $c_2=-\frac{1}{4}{\cal S}$ and $c_4=\frac{1}{4}{\cal T}$ of Eqs.(\ref{eh1})-(\ref{eh2}).

As a solid piece of evidence it is also obvious that $P_2=\langle
{\cal Q}_1,{\cal Q}_2\rangle$ with $\langle,\rangle$ being the
usual symplectic product of charge vectors which is a singlet with
respect to $SL(2)_0\times G_4$. Hence for mutually local charge
configurations our cubic curve of Eq.(\ref{magiccurve}) falls
short of the term proportional to $xy$ just like in the $stu$
truncation. Moreover, as shown by
Eqs.(\ref{fourvectorsdual})-(\ref{26})  in the four-qubit case our
quadratic and sextic invariants $I_1$ and $I_3$ are duals of each
other . Hence we expect that the invariant $P_6$ should be related
to a quadratic combination of the Freudenthal duals of the
corresponding charge vectors which are cubic in terms of the
original charges\cite{fdual}. Moreover, these dual quantities
should be antisymmetric with respect to the exchange of the
centers. Luckily a quantity $J_6$, satisfying these criteria is
also available (see Eq. (3.23) of Andrianopoli et.al.\cite{Ferretal2}).

Unfortunately since according to Eq.(\ref{relatedinv})
$D=I_3-I_1I_4$ in the $stu$ case this quantity is probably not
directly related to our invariant $P_6$. Hence one is still left
with the problem of finding the invariants $P_4,P_8$ and $P_{12}$.

 In order to gain some insight into the physical meaning
of the invariant $P_4$ let us look again at the $stu$ (four-qubit)
case. In this case we know that the corresponding invariant is
$M-L$. As a first step we would like to somehow relate this
quantity to some invariant known within the context of $N=8$,
$d=4$ supergravity.

For the $stu$ truncation let us suppose that we have chosen a
particular two-center charge configuration parametrized by the
four real numbers $a,b,c,d$ of the canonical form of
Eq.(\ref{genuine}). Now the explicit form of the invariants
$H,M,L,D$ is

\beq H=\frac{1}{2}(a^2+b^2+c^2+d^2),\quad
M=\left[\left(\frac{c-d}{2}\right)^2-\left(\frac{a-b}{2}\right)^2\right]
\left[\left(\frac{a+b}{2}\right)^2-\left(\frac{c+d}{2}\right)^2\right]
\eeq \noindent

\beq D=\frac{1}{4}(ad-bc)(cd-ab)(ac-bd).\qquad
L=abcd.\eeq\noindent 

Consider now the quartic $E_{7(7)}$ invariant\cite{Becker}
expressed in terms of the matrix of central charge ${\cal Z}$ of
$N=8$ supergravity in $d=4$. This matrix is an $8\times 8$ complex
antisymmetric one which can be brought to the canonical form after
using a suitable
 $U(8)$ transformation ${\cal Z}\mapsto U^T{\cal Z}U$.
This canonical form is \beq U^T{\cal
Z}U=\begin{pmatrix}z_0&0&0&0\\0&z_1&0&0\\0&0&z_2&0\\0&0&0&z_3\label{kanform}\end{pmatrix}\otimes
\begin{pmatrix}0&1\\-1&0\end{pmatrix}\eeq\noindent
 Now Cartan's quartic invariant is \beq J_4({\cal Z})\equiv {\rm
Tr}({\cal Z}\overline{{\cal Z}}{\cal Z}\overline{{\cal
Z}})-\frac{1}{4}({\rm Tr}({\cal Z}\overline{{\cal Z}}))^2+4({\rm
Pf}({\cal Z})+{\rm Pf}(\overline{{\cal Z}}))\eeq\noindent where
the Pfaffian is\beq {\rm Pf}({\cal
Z})=\frac{1}{2^44!}{\varepsilon}^{ABCDEFGH}{\cal Z}_{AB}{\cal
Z}_{CD}{\cal Z}_{EF}{\cal Z}_{GH},\qquad A,B,\dots=1,2,\dots
8\eeq\noindent Subscripts $A,B\dots$ label an of ${\bf 8}$
$SU(8)$. Using the canonical form one gets the well-known
expression \beq J_4 =(\vert z_0\vert^2+\vert z_1\vert^2+\vert
z_2\vert^2+\vert z_3\vert^2)^2-4(\sum_{n<m}\vert
z_nz_m\vert^2)+8{\rm Re}(z_0z_1z_2z_3) \eeq\noindent It is also
known that by an $SU(8)$ transformation it is possible to remove
three phases, so for instance $z_1,z_2,z_3$ can be chosen to be
real (or else having the same phase). Hence the canonical form can
be characterized by four real numbers $r_n=\vert z_n\vert$ and a
phase $\phi$. The form of $J_4$ reflecting these considerations is
\begin{eqnarray}
J_4&=&[(r_0+r_1)^2-(r_2+r_3)^2][(r_0-r_1)^2-(r_1-r_2)^2]+8r_0r_1r_2r_3(\cos\phi-1)\nonumber\\
&=&-16M+8L(\cos\phi-1)=8(N-M)+8L\cos\phi\nonumber
\end{eqnarray}\noindent
where we have used the identity $L+M+N=0$ of Eq.(\ref{osszef}) and
we made the identifications $r_0=a, r_1=b,r_2=c$ and $r_3=d$.
Hence though we did not manage to get directly to the invariant
$M-L$ however for $\phi=\frac{\pi}{2}$ we get $N-M$ instead which
is related to $M-L$ via a permutation of the qubits labelled by
$i_1i_2i_3$ carrying the $G_4$ labels. Notice also that the
permutation group $S_3$ responsible for this triality symmetry at
the $stu$ model level is related to triality of the complex group
$SO(8,{\mathbb C})$ whose real form $SO(4,4)$ is featuring the tripartite
entanglement of seven qubits interpretation of Cartan's quartic
invariant\cite{ferfano,fanolevay} of $N=8$ $d=4$ supergravity and
the corresponding $stu$ truncation. Notice also that according to
Eq.(\ref{kellenifog})  $M-N$ is just the invariant $\chi$ of
Ferrara et.al.\cite{Ferretal}. These considerations show that the
unknown invariant $P_4$ should truncate to $\chi=N-M$  after some
suitable permutations.

It is also interesting to notice that for $z_n$ {\it real} i.e.
$\phi=0$ we have \beq
J_4=-16M=16\tilde{r}_0\tilde{r}_1\tilde{r}_2\tilde{r}_3.\eeq\noindent
 Here
 \beq \tilde{r}_n=\sum_{m=0}^3(H\otimes
H)_{nm}r_m, \qquad
\begin{pmatrix}\tilde{r}_0\\\tilde{r}_1\\\tilde{r}_2\\\tilde{r}_3\end{pmatrix}=\frac{1}{2}\begin{pmatrix}1&1&1&1\\1&-1&1&-1\\
1&1&-1&-1\\1&-1&-1&1\end{pmatrix}\begin{pmatrix}r_0\\r_1\\r_2\\r_3\label{hada2}\end{pmatrix}\eeq\noindent
where $H$ is the Hadamard matrix used for implementing discrete
Fourier transformation in quantum information theory. Notice that
with the definition $Q_n=2\tilde{r}_n$ we have $J_4=Q_0Q_1Q_2Q_3$
the standard expression giving rise to the Bekenstein-Hawking
entropy $S=\pi\sqrt{4Q_0Q_1Q_2Q_3}$ of the $D2-D2-D2-D6$ brane
configuration in the type IIA duality frame\cite{Becker}.

There still remained the task to understand the meaning of $P_{8}$
and $P_{12}$. Since in the $N=2$ , $d=4$ context the vanishing of
the corresponding invariants identifies the $stu$, $st^2$ and
$t^3$ truncations in terms of the structure of the
(\ref{magiccurve}) elliptic curve it would be important to
know their explicit forms. Finally the discriminant ${\cal
S}^3-27{\cal T}^2$ as a polynomial invariant of order $24$ should
play the role of some sort of generalization of the
hyperdeterminant of type $2\times 2\times 2\times2$. It would be
interesting to clarify what kind of role this discriminant and the
associated $j$ function plays in the physics of two-centered black
hole solutions.

\section{A triality symmetric curve}

The aim of this speculative section is to draw the readers attention to some interesting structural similarities showing up in a variety of physical contexts where our four-qubit invariants parametrizing elliptic curves might play a crucial role.

In the previous investigations our basic philosophy for the association of elliptic curves to
two-center charge configurations was based on the resolvent cubic
of the fundamental polynomial of Eq.(\ref{poli1}).
Is there any
{\it other} physically appealing way for this association which
retains the fundamental role of this polynomial and at the same time also featuring
an extra modular parameter $\hat{\tau}$ accounting for a hidden torus? In order to show that the
answer to this question is yes let us rewrite Eq.(\ref{poli1}) in
the form

\beq
\Sigma_4(\Lambda_{i_0i_1i_2i_3},\hat{\tau},-x)=x^4+(\alpha\beta)(4I_1)x^3+(\alpha\beta)^2(6I_2)x^2
+(\alpha\beta)^3(4I_3)x+(\alpha\beta)^4I_4^2.\label{deformedsigma}\eeq\noindent The
quantities displaying explicit dependence on $\hat{\tau}$ are
defined as \beq \alpha=e_2-e_1,\qquad \beta=e_3-e_1 \eeq\noindent
where

\beq 4(x-e_1)(x-e_2)(x-e_3)=4x^3-g_2x-g_3 \eeq\noindent with $g_2$
and $g_3$ given by Eqs.(\ref{eisenstein})-(\ref{eisenstein2}).
Explicitly we have\cite{Kra,SW}
 \beq e_1-e_2=\theta_3^4(0,\hat{\tau}),\qquad
e_3-e_2=\theta_1^4(0,\hat{\tau}),\qquad
e_1-e_3=\theta_2^4(0,\hat{\tau})\eeq\noindent

where \beq \theta_1(0,\hat{\tau})=\sum_{n\in{\mathbb
Z}}q^{\frac{1}{2}(n+1/2)^2}\eeq\noindent

\beq \theta_2(0,\hat{\tau})=\sum_{n\in{\mathbb
Z}}(-1)^nq^{\frac{1}{2}n^2}\eeq\noindent

\beq \theta_3(0,\hat{\tau})=\sum_{n\in{\mathbb
Z}}q^{\frac{1}{2}n^2}\eeq\noindent with $q=e^{2\pi i\hat{\tau}}$.
One also has\cite{Kra,SW} \beq
e_1=\frac{2}{3}+16q+16q^2+\dots\eeq\noindent \beq
e_2=-\frac{1}{3}-8q^{\frac{1}{2}}-8q-32q^{\frac{3}{2}}-8q^2+\dots\eeq\noindent
\beq
e_3=-\frac{1}{3}+8q^{\frac{1}{2}}-8q+32q^{\frac{3}{2}}-8q^2+\dots\eeq\noindent
Notice that $e_1+e_2+e_3=0$ similarly to the property of four
qubit quartic invariants $L+M+N=0$. In the limit $\hat{\tau}\to
i\infty$ we have $\alpha\beta\to 1$ hence from Eq. (\ref{deformedsigma}) we get back to our usual
polynomial of Eq.(\ref{poli1}). Clearly
introducing the factor $\alpha\beta$ amounts to rescaling
the amplitudes of our four-qubit state of Eq.(\ref{Lambda})
from $\Lambda$ to the $\hat{\tau}$ dependent ones
$\sqrt{\alpha\beta}\Lambda$.

Let us now introduce an extra complex parameter $u$ to be specified later and consider
the following family of elliptic curves \begin{eqnarray} y^2= x(x-\alpha
u)(x-\beta u)&-&
(\alpha-\beta)^2I_1x^2+\frac{1}{2}(\alpha-\beta)\alpha\beta[(\alpha+\beta)I_4-3(\alpha-\beta)I_2]x
\nonumber\\
&-&(\alpha-\beta)\alpha^2\beta^2I_4u-(\alpha-\beta)^2\alpha^2\beta^2I_3.\label{SWC1}
\end{eqnarray}\noindent
Now the right hand side is a quadratic polynomial in $u$. Its
discriminant turns out to be just
$(\alpha-\beta)^2\Sigma_4(\Lambda,\hat{\tau},-x)$.
This gives the desired clue for yet another way of associating an elliptic curve to two-center charge configurations.

Notice that in the limit $\hat{\tau}\to i\infty$ our curve boils
down to $y^2=x(x-u)^2$ a curve similar to the one of Eq. (\ref{nbpscurve}) with $j$ invariant $\infty$. Moreover, via
the transformation $v=u+(\alpha -\beta)\frac{1}{2}H$ one can also
obtain a new form
\begin{eqnarray} y^2&=& x(x-\alpha v)(x-\beta
v)+\beta(\alpha-\beta)Hx^2-\alpha\beta(\alpha -\beta)(\alpha M
+\beta N)x\nonumber\\&-&\alpha\beta(\alpha-\beta)Hvx
-\alpha^2\beta^2(\alpha -\beta)Lv-\alpha^2\beta^2(\alpha
-\beta)^2D.\label{newstu}\end{eqnarray}\noindent Though not in the Tate form
this curve is displaying similar quantities than our previous
curve of Eq.(\ref{stucurve}).
In particular for truncations from $stu$ charge configurations to $st^2$ and $t^3$ ones the structure of the curve is getting simpler step by step.
Though this curve now shows an explicit dependence on a modular
parameter $\hat{\tau}$ and also featuring a new complex variable $v$, now its discriminant cannot obviously be related to a hyperdeterminant of type $2\times 2\times 2\times 2$.
However, curves like Eq.(\ref{newstu}) have other appealing properties which we would like to discuss.

Actually the curve of Eq.(\ref{SWC1}) is related to a 
one introduced by Seiberg and Witten in their study of $N=2$
supersymmetric $SU(2)$ gauge theory with four quark
flavours\cite{SW}. The parameter $u$ in that case was the gauge
invariant modulus representing the square of the Higgs expectation
value and $\hat{\tau}$ was the complex coupling constant
$\theta/\pi+8\pi i/g^2$ of Montonen-Olive duality. In that context
their curve was parametrized not by algebraically independent
four-qubit invariants but by the squares of the four quark masses
$m_1,m_2,m_3$ and $m_4$. However, after comparing the relevant
expressions (see in particular Eq. 17.58 of Ref.\cite{SW} )  it is
easy to show that the quark masses squared in that case correspond
to the squares of the parameters $a,b,c$ and $d$ of our four-qubit
canonical form of Eqs.(\ref{genuine})-(\ref{genpar}).

 Apart from the symmetry
$SL(2,{\mathbb R})_0\times G_4$ where $G_4=SL(2,{\mathbb
R})^{\times 3}$ is the continuous $U$-duality group in the
supergravity approximation, the $stu$ model also has an important
triality symmetry\cite{stualap}. In our four-qubit description
this $S_3$ permutation symmetry should manifest itself in
representing somehow elliptic curves in a way displaying four
qubit invariants that are also invariant under permutation. In
analogy with the Seiberg-Witten curve  there is also the possibility to present
a triality invariant form of that kind.
For this purpose we have to
chose from four algebraically independent four-qubit invariants
which are also invariant under the full permutation group $S_4$.
Such invariants were first constructed by Schl\"afli\cite{Schl} in
1852. Let us denote these invariants as\cite{Djokovic,Osterloh}
$H,\Gamma,\Sigma$ and $\Pi$. They are of order $2,6,8$ and  $12$
respectively. The new quantities $\Gamma,\Sigma$ and $\Pi$ are
defined as \beq \Gamma =D+E+F,\qquad \Sigma=L^2+M^2+N^2,\qquad \Pi
=(L-M)(M-N)(N-L). \eeq\noindent Here $E$ and $F$ are defined
as\cite{Luque} \beq HL=E-D,\qquad HM=D-F,\qquad
HN=F-E.\eeq\noindent Notice in particular that in terms of the
quantities $H,L,M,D$ used to describe the Tate form of the
elliptic curve of Eq.(\ref{stucurve}) the permutation invariant
version of the sextic invariant reads as \beq
\Gamma=3D+H(L-M).\label{perminv6}\eeq\noindent

Let us now consider the triality symmetric curve\cite{SW}
\beq
y^2=w_1w_2w_3+\frac{\lambda}{3}[(M-N)(e_2-e_3)w_1+(N-L)(e_3-e_1)w_2+(L-M)(e_1-e_2)w_3]-
\frac{\lambda^2}{3}\Gamma\label{swcurve}\eeq\noindent where \beq
w_i=x-e_i\tilde{u}-e_i^2H,\qquad
\lambda=(e_1-e_2)(e_2-e_3)(e_3-e_1)\qquad i=1,2,3.\eeq\noindent
Notice that after using Eq.(\ref{perminv6}) and the canonical form
for our four qubit states of Eq.(\ref{genuine}) and (\ref{gyokok})
with $(t_1,t_2,t_3,t_4)=(a^2,b^2,c^2,d^2)$ for the polynomial
$\Gamma$ of sixth order we obtain the expression \beq
\frac{1}{3}\Gamma=\frac{3}{16}\sum_{i>j>k}t_it_jt_k-\frac{1}{96}
\sum_{i\neq j}t_it_j^2+\frac{1}{96}\sum_it_i^3.\eeq\noindent This
is precisely the sixth order invariant of Eq. (16.36) of Seiberg
and Witten\cite{SW} provided we make the identification $t_i\equiv
m_i^2$ i.e. the canonical four-qubit parameters are identified with
the quark masses of the four flavours. Similar calculations verify
that the remaining invariants of that paper namely $R,T_1,T_2$ and
$T_3$ can be identified with the invariants
$H,\frac{1}{3}(M-N),\frac{1}{3}(N-L)$ and $\frac{1}{3}(L-M)$ in
the canonical parametrization. The important property of the curve
of Eq.(\ref{swcurve}) is that after making the transformations
\beq u=\tilde{u}+\frac{1}{2}e_1H,\qquad x\to
x-\frac{1}{2}e_1u+\frac{1}{2}e_i^2H\eeq \noindent and performing
the (weak coupling\cite{SW}) limit $\hat{\tau}\to i\infty$ it
boils down to the form \beq
y^2=x^2(x-u).\label{osszevet}\eeq\noindent
Moreover, in order to make contact with our original curve of Eq.(\ref{SWC1})
one just has to perform the change of variables in Eq.(\ref{swcurve})
\beq
x\mapsto x-e_1\tilde{u}-e_1^2H,\qquad u=\tilde{u}+\frac{1}{2}e_1H.\eeq\noindent

It is also interesting to realize that our identification of the quark masses squared with the canonical four-qubit parameters automatically incorporates the special cases when some of the masses are zero with the singular cases in the four-qubit classification scheme of Verstraete\cite{Verstraete}.
Moreover, switching to the two-center $stu$ context there the canonical parameters would rather be identified with the the parameters $z_1,z_2,z_3$ and $z_4$ of the canonical form of the central charge matrix of $N=8$ $d=4$ supergravity (see also the discussion of the previous seubsection).
Recall also in this respect the observation of Seiberg and Witten\cite{SW} that triality symmetry of $SO(8)$ is connected to the permutation symmetry $S_3$ as the mod 2 reduction of $SL(2,{\mathbb Z})$.
Comparing Eqs. (17.34)-(17.35) of Ref.14 with our Eq.(\ref{hada2})
related to the structure of the matrix ${\cal M}$ and a similar expression related to the matrix ${\cal N}$
that are in turn related to the structure of four-qubit reduced density matrices(see Eqs.(\ref{redsur})-(\ref{redsur2})) we see that triality symmetry is intrinsically related not only to the structure of four-qubit entanglement but to the structure of the manifold $X$ described by the curve of Eq.(\ref{swcurve}).
In this context it is especially instructive to recall the arguments of Seiberg and Witten  on the structure of the cohomology of $X$ . In particular it is tempting to reinterpret the triality invariant expressions for the cohomology classes of the periods\cite{SW} as genuine unnormalized four-qubit states in canonical form (see Eqs. 17.36 and 17.37 of that paper). The redundancy in arriving at such an expression can be attributed to the action of the Weyl group of $SO(8)$ on the canonical parameters (i.e. the action of the Klein group on the canonical form of $\vert\Lambda$) in accord with a comment
of Luque and Thibon in the four-qubit context\cite{Luque}.
(See also the last paragraph of Section II. in this respect.)

Finally it is amusing to recall yet another context where elliptic
curves parametrized by four-qubit invariants reveal some
intriguing structural similarities with interesting physics. First
of all it is well-known that in F-theory an extra $SL(2,{\mathbb
Z})$ and a hidden torus also makes its presence in a spectacular
way\cite{Becker,Vafa}. In this case one considers an elliptic
fibration $M$ with some basis manifold $B$ and fiber a two
dimensional torus with modular parameter $\hat{\tau}$. Now
$F$-theory is defined on $M$ as type $IIB$ string theory on $B$
with the axion dilaton modulus of type $IIB$ string theory
identified with the modular parameter of the two-torus. In the
original setting\cite{Vafa} an elliptically fibered $K3$ surface
was considered of the form $ y^2=x^3+f(u)x+g(u)$ where $f$ and $g$
are polynomials of degree eight and degree twelve in $u\in
{\mathbb C}P^1$. This curve describes a torus for each point of
the Riemann sphere ${\mathbb C}P^1$ labelled by the complex
coordinate $u$. It is interesting to realize that this
elliptically fibered $K3$ surface is of the form of Eq.
(\ref{kifuto}) where $S$ and $T$ are polynomials of order eight
and twelve. However, according to Eq.(\ref{genuine}) when writing
$S$ and $T$ in the canonical form we are having four complex
coordinates instead of the one $u$. The points where the torus
degenerates corresponds to the vanishing of the discriminant of
the cubic which is a polynomial of order $24$. In the four qubit
parametrization this just corresponds to the vanishing of our
hyperdeterminant of order $24$ related to  the quantity
$S^3-27T^2$. In the original $F$-theory context the
compactification described by this elliptical fibration
corresponds to a configuration of $24$ seven branes of type IIB
theory located at the zeros of the discriminant. On the other hand
in the four-qubit canonical parametrization the explicit
expression for the discriminant of Eq.(\ref{gyokok}) is similar to
the usual ones describing coincident seven brane
configurations\cite{senagain}, provided we are regarding one of the canonical parameters to be special.

Notice also that the study of $F$-theory
compactifications\cite{Collinucci} is effected by looking at
elliptic fibrations described by elliptic curves of the Tate form
similar to the one of Eq.(\ref{magiccurve}) where now the
coefficients are polynomials in the coordinates of the base
manifold $B$. An important limit studied in these investigations
is the Sen limit\cite{senagain} or orientifold limit. This is
achieved by demanding that the axio-dilaton to be constant almost
everywhere in type IIB space-time. Explicitly this limit is
realized by setting \beq a_3\mapsto \varepsilon a_3,\qquad
a_4\mapsto \varepsilon a_4, \qquad a_6\mapsto
\varepsilon^2a_6,\eeq\noindent in the Tate-form. In the two center
$stu$-context $a_i$ are polynomial invariants of homogneous degree
$2i$ and $a_6=0$. For the $stu$ model we have $a_4=-LM$, $a_3=D$.
For the $st^2$ and $t^3$ models $L=0$ hence $I_3=D$ and a
corresponding limit is achieved by sending the invariant $D$ to
zero by e.g. scaling down one of the charges to zero. This is
precisely what happened in Section V.B. and also in the situation
described by Eq.(\ref{j0}).

Finally our triality symmetric Seiberg-Witten curve of
Eq.(\ref{swcurve}) also makes its presence in the $F$-theory
context\cite{senorient}. Here the basic idea is to deform away
from the special point in moduli space where the orientifold
picture applies. This important deformation is effected by a curve
of the form $y^2=x^3+\tilde{f}(u)x+\tilde{g}(u)$ where $\tilde{f}$
and $\tilde{g}$ are polynomials in $u$ of degree two and three.
Sen has shown that this curve can be cast in the form of
Eq.(\ref{swcurve}) provided we make a suitable mapping of the
parameters involved in the two different physical contexts. Notice
that in the original papers the relevant curves\cite{SW,senorient}
were not parametrized by the four-qubit invariants $H,L,M,N$ and
$\Gamma$  as in Eq.(\ref{swcurve}). Instead in these papers four
parameters were employed, which are easily identified with the
four canonical parameters of four-qubit states of
Eq.(\ref{genuine}). In particular in the F-theory context the
relevant deformation is a one corresponding to splitting of the
six coincident zeros of the discriminant away from each other, i.e
in the orientifold picture moving the four coincident seven branes
away from the orientifold plane. The collective coordinates of the
background describing such a physical situation are that of an
$N=1$ supersymmetric $SO(8)$ gauge theory in eight dimensions with
moduli space characterized by a complex scalar field ${\Phi}$ . At
a generic point in moduli space the vacuum expectation value of
this field has the form similar to the canonical form of the
central charge of Eq.(\ref{kanform}) we used in Section V. in the
different context of $N=8$ supergravity. Now one can check that
the four canonical complex parameters $c_1,c_2,c_3$ and $c_4$ of
$\langle\Phi\rangle$ are behaving exactly like the canonical
parameters of a four-qubit state. Again the polynomial expressions
used by Seiberg and Witten in their curve of the
Eq.(\ref{swcurve}) form are the ones for the four-qubit invariants
$H,L,M,N$ and $\Gamma$ now with the simple
identification\cite{senorient} $c_i=m_i$. Our four-qubit analysis
as presented in this paper shows similar symmetry properties
(permutation symmetry of the stu truncation  connected to triality
of $SO(8)$ within $N=8$ supergravity ). This might indicate that
the physics of two-center black holes could be another arena
where this curve plays a basic role. Of course these  structural
similarities could be superficial however, in any case the possible
physical ramifications should be explored further.

\section{Conclusions}

In this paper we have shown how the $U$-duality invariants
introduced by Ferrara et.al.\cite{Ferretal} characterizing
two-center extremal black hole charge configurations in the $stu$,
 $st^2$ and $t^3$ models of $N=2$, $d=4$ supergravity can be
understood as entanglement invariants of four-qubit systems. In
this entanglement based picture the geometric and algebraic
meaning of these invariants is displayed in a nice and unified
manner. For one of the entanglement invariants that have not yet
made its debut to the supergravity literature we have found a
distinguished role. It is the hyperdeterminant of type $2\times
2\times 2\times 2$ which is the generalization of Cayley's
hyperdeterminant featuring the macroscopic black hole entropy
formula in the $stu$ model. For the special example of the BPS
$D0D4-D2D6$ split in the $t^3$ model we have demonstrated that
this polynomial invariant of order $24$ seems to govern important
issues of consistency for the two center solutions.

We have also introduced a quartic polynomial featuring the
algebraically independent four-qubit invariants. For our simple
example we have shown that the property that this polynomial has
real roots provides a necessary condition for the consistency
condition to hold. The resolvent cubic of this fundamental
polynomial can be cast into a cubic of Weierstrass canonical form
which in turn can be used to define an elliptic curve associated
to the two-center charge configuration. After switching to the
more convenient Tate form of this curve we have shown that this
association is natural, meaning that the Tate form is displaying
combinations of the algebraically independent invariants of
physical meaning. Indeed, the Tate form falls short of terms step
by step as we perform truncations to the $st^2$ and $t^3$ models,
restrict attention to mutually local charge configurations, or
perform the degenerate limit resulting in splits into small black
holes. The discriminant of this elliptic curve is just the
hyperdeterminant, a quantity also featuring the $j$ invariant of
the curve. For our example the structure of the $j$ invariant
nicely encapsulates the basic properties of the $D0D4-D2D6$ split.

The mapping from two center black hole charge configurations to
elliptic curves is many to one. Moreover, for the $stu$ model only
a special class of curves with $a_6=0$ in Eq.(\ref{stucurve}) can
be reached. We observed that the vanishing of $a_6$ can be
attributed to the vanishing of a polynomial of degree $12$
characterizing the $stu$ model. Based on this we conjectured that
our $N=2$ picture can be substantially generalized also
incorporating two center charge configurations of $N=8$
supergravity. Here Cartan's quartic invariant and its
polarizations should make their presence together with the usual
symplectic invariant and a new invariant of order $6$ introduced
recently\cite{Ferretal2}. The discriminant of this curve of more
general type then would play the role of a generalization of our
hyperdeterminant, with probably similar physical meaning than its
$stu$ descendent.

Finally we presented other physical contexts where our four-qubit
invariants parametrizing elliptic curves also play an important
role. Here we have given a new look to the triality invariant
curve originally introduced by Seiberg and Witten\cite{SW}. In its
new form this curve is featuring four-qubit invariants also
displaying permutation invariance. In this new setting we have
also invoked the F-theory interpretation of this curve as was
given by Sen\cite{senorient}. Taken together with the two-center
black hole context studied in this paper the main unifying theme
in these scenarios seems to be the presence of a hidden torus and
its extra set of $PSL(2,{\mathbb Z})$ modular transformations.
Note that in our investigations elliptic curves appeared merely as
natural mathematical objects nicely encapsulating the information
on the structure of two-center $U$-duality invariants. However,
apart from this we have also presented some evidence that the
hyperdeterminant and the $j$ invariant associated to these curves
might contain useful information on issues of marginal stability,
domain walls and split attractor flows. The physical background
underlying the use of elliptic curves in the F-theory and $N=2$
supersymmetric $SU(2)$ gauge theory context is well-known. Is
there any deeper physical reason also accounting for the
occurrence of elliptic curves related to the structure of
multi-center black hole solutions?

\section{Acknowledgement}
This work was supported by the New Hungary Development Plan
(Project ID: T\'AMOP-4.2.1/B-09/1/KMR-2010-0002).

\end{document}